\documentclass[preprints,article,accept,oneauthor]{Definitions/mdpi} 
\def\ie{{\em i.e., }}
\def\be{\begin{equation}}
\def\bea{\begin{eqnarray}}
\def\eea{\end{eqnarray}}
\def\ee{\end{equation}}
\firstpage{1} 
\pubvolume{1}
\issuenum{1}
\articlenumber{0}
\pubyear{2022}
\copyrightyear{2022}
\datereceived{} 
\dateaccepted{} 
\datepublished{} 
\hreflink{https://doi.org/} % If needed use \linebreak
\Title{Effect of fluid composition on a jet breaking out of a cocoon in gamma-ray bursts : A relativistic de Laval nozzle treatment}

\TitleCitation{Title}

\Author{Mukesh K Vyas $^{\dagger}$\orcidA{}}%, Firstname Lastname $^{2,\ddagger}$ and Firstname Lastname $^{2,}$*}

\AuthorNames{Firstname Lastname, Firstname Lastname and Firstname Lastname}

\AuthorCitation{Lastname, F.; Lastname, F.; Lastname, F.}
\address{%
$^{\dagger}$ \quad Department of Physics, Bar Ilan University, Ramat Gan, Israel (5290002); mukeshkvys@gmail.com}%\\
\corres{Correspondence: mukeshkvys@gmail.com}

\abstract{We carry out a semi-analytic general relativistic study of a Gamma-Ray Bursts (GRB) jet that is breaking out of a cocoon or stellar envelope. We solve hydrodynamic equations with the relativistic equation of state that takes care of fluid composition. In short GRBs, a general relativistic approach is required to account for curved spacetime in strong gravity. The piercing of the jet through the cocoon resembles a de Laval nozzle and the jet may go through recollimation shock transitions. We show that the possibility of shock transition and the shock properties are sensitive to the matter composition and the cocoon strength. The obtained Lorentz factors in thermally driven jets comfortably reach a few $\times~10$.
}


\begin{document}

%%%%%%%%%%%%%%%%%%%%%%%%%%%%%%%%%%%%%%%%%%
%\setcounter{section}{-1} %% Remove this when starting to work on the template.
%\section{How to Use this Template}
%
%The template details the sections that can be used in a manuscript. Note that the order and names of article sections may differ from the requirements of the journal (e.g., the positioning of the Materials and Methods section). Please check the instructions on the authors' page of the journal to verify the correct order and names. For any questions, please contact the editorial office of the journal or support@mdpi.com. For LaTeX-related questions please contact latex@mdpi.com.%\endnote{This is an endnote.} % To use endnotes, please un-comment \printendnotes below (before References). Only journal Laws uses \footnote.

% The order of the section titles is: Introduction, Materials and Methods, Results, Discussion, Conclusions for these journals: aerospace,algorithms,antibodies,antioxidants,atmosphere,axioms,biomedicines,carbon,crystals,designs,diagnostics,environments,fermentation,fluids,forests,fractalfract,informatics,information,inventions,jfmk,jrfm,lubricants,neonatalscreening,neuroglia,particles,pharmaceutics,polymers,processes,technologies,viruses,vision

\section{Introduction}
In the current understanding of gamma-ray bursts (GRBs), short GRBs are produced when two compact objects merge together \cite{npp92,je99,fwh99}, while a long GRB is a result of the collapse of a massive star \cite{le93, w93, mw93}. In both cases, a bipolar jet is launched from the interior of the star or merger and escapes to produce the observed signatures. While breaking out of the stellar surface, the jet interacts with a dense medium or stellar envelope ahead of it. The interaction leads to the deceleration of the jet and the formation of shocked matter at the jet-envelope interface. This piled up matter is an extended structure known as a stellar cocoon or cork. The cocoon then warps up the jet and affects its dynamic evolution \cite{ma09}.
% which is either the outer shell of the star or produced by the piled up matter due to the jet head. 
The properties of the cork are modeled extensively in the literature \cite{bc89,m03, bnp11, nhs14,bmr14,ldm16, mrm17, hgn18, gnp18, gnph18, yb22}. This cocoon forms in both the short and long GRBs \cite{np16}. Depending upon the kinetic power of the jet, it either pierces through the cocoon to escape to the external medium \cite{rml02,zwm03,zwh04}, or it is choked to produce backscattered photons \cite{em08,e14,e18,vpe21a,vpe21b, vpe21c}. In the former case, the escaped jet's geometry is shaped and its dynamics is significantly affected by the jet cork interaction \cite{nhs14,hgn18, m03,mlb07,ma09,mi13, dqk18, gln19, hi21, gnb21,gg21}. The cocoon is capable of inducing recollimation shocks in the jet stem \cite{bnp11,gln19}. To explain the observed features of the bursts, the formation of a cocoon is found to be an essential requirement. In recently discovered short GRB associated with gravitational source GW170817, the radio observations hint towards the presence of a cocoon as the jet erupts from the burst \cite{hcm17,mnh18,gbs20}. It is found that the cocoon shock breakout of the jet in the case of GW170817 is a more likely picture over other scenarios such as photospheric emission \cite{gnph18,ijt21}. 

Given the fact that the jet interacts with the cocoon before breaking out of the system, one may ask an obvious question about how the strength of the cocoon and the composition of the jet matter affect its dynamical evolution during this process. The jet composition is an open question in GRB physics. It is debatable whether GRBs are matter-dominated or radiation dominated \cite{mkp06,kz15,p15, cg16, fcz17, zzc18,cpd21,gkg21,zwl21}. In this study, we consider a jet composed of baryons and leptons and we keep the lepton fraction over baryons to be a free parameter, an approach similar to \cite{zwl21}. It is noted that the neutrino annihilation in the initial phase of the burst leads to the formation of a large amount of electron-positron pairs \cite{w93,pwf99,mw99,le03,gl14} and thus the positrons are likely to be a part of the outflowing jet. In the general relativistic regime, we aim to investigate further properties of the jet once it interacts and is shaped by the cocoon. We pose and seek to answer the following questions in this analysis: \begin{itemize}
%\item Considering that the jet is composed of matter, how its dynamics is affected by its interaction with the cocoon;\\
\item As the jets in short GRBs produce near strong gravitational potential, how the jet scenario is affected in the regime of general relativity;
\item Once the jet strikes against the cocoon to be collimated and escaped, how its dynamics is sensitive to its matter composition (\ie ratio of lepton's fraction to baryonic matter) and how its observational appearance changes.;
\item What are the conditions for the generation of the recollimation shocks produced by collimation of the jet and how the possibility of the shock transition depends upon matter composition?
\item How the properties of the shock (like shock strength and transition location) are sensitive to the jet composition?
%\item 
\end{itemize}
Hence, we project to carry out an extensive analysis of the dependence of jet dynamics on its composition, as well as, the collimation strength of the cocoon. To meet this purpose, we solve general relativistic hydrodynamic equations of motion using a relativistic equation of state proposed by \citep{cr09}. This equation of state takes care of the fluid composition. The jet matter made of electrons and protons may harbour a significant fraction of positrons as pair production takes place at high optical depths \cite{gcs08}.

In the next section, we describe the assumptions in the model in detail including the considered jet geometry and the equation of state. Then we proceed to discuss the jet dynamics along with the equations of motion in the general relativistic regime in section \ref{sec_dyn}. In section \ref{sec_method}, we discuss the method to solve the dynamical equations of motion using sonic point analysis and discuss the shock conditions and shock properties. The results are described in section \ref{sec_results}. We conclude the paper in section \ref{sec_conclusions} to highlight the principle outcomes and their significance in understanding the GRBs. In brief, we discuss the future extension of this work in section \ref{sec_future}.
\section{Assumptions}
%Jet has mechanical interaction with cocoon, one d study, metric, steady state jet
In this work, we construct a semi-analytic general relativistic steady-state model of an erupting jet which is shaped and collimated by the stellar cork. Although the cocoon forms both in long and short GRBs, we restrict this paper to the case of a short GRB jet erupting from the merger of two neutron stars. The conclusions should have morphological similarities with the long GRB jets. However, general relativistic consideration is necessary for short GRBs only. As the jet is hot and harbours relativistic temperatures, one needs a relativistic equation of state to account for the variable nature of the adiabatic index with its temperature. We describe the equation of state in section \ref{sec_eos} below. 

A typical neutron star has mass $1.4-3.6 M_\odot$ and a radius $10-12$ Km or roughly $1.5-3$ Schwarzschild radius ($r_g$) \cite{nc73,rr74,kb96,st08}. When two neutron stars merge, we expect that the jet should be erupting from the surface of the merger.  It is clear that the analysis needs the inclusion of curved space-time. We consider the Schwarzschild metric and work with geometric units where length and time are defined in the units $GM/c^2$ (or $r_g/2$) and $r_g/2c$, respectively. Here $G$ is the universal constant of gravity, $M$ is the total mass after the merger and $c$ is light speed. Hence, the velocity is defined in terms of $c$. The base of the jet is considered at $R^*=2 r_g$, or we treat it as the surface of the star.

After launching, the jet interacts mechanically with the envelope above the stellar surface and is shaped accordingly. Properties of the considered jet cross-section are inspired by various numerical studies and are described in detail in section \ref{sec_cross} below. It is a one-dimensional study, \ie we consider that the jet's local properties are constant across its horizontal width. This assumption allows us to treat the problem in one dimension and it is justified for narrow jets with a small opening angle which is reasonable for GRBs. We treat the cocoon as an auxiliary agent that shapes the jet and defines its geometry and do not consider any energy exchange between both. It means that in the adiabatic equation of state, the jet energy parameter remains constant and is not affected by its interaction with the cork but its dynamical evolution changes. Further, as the jet speed is likely to be significantly greater than the outflowing speed of the cocoon, we simply consider a stationary cross-section of the jet. It implies that the cocoon shapes the jet but its collimation effects don't evolve with time.
\label{sec_assump}

\begin{figure}[H] \begin{center}
\includegraphics[width=10.5 cm]{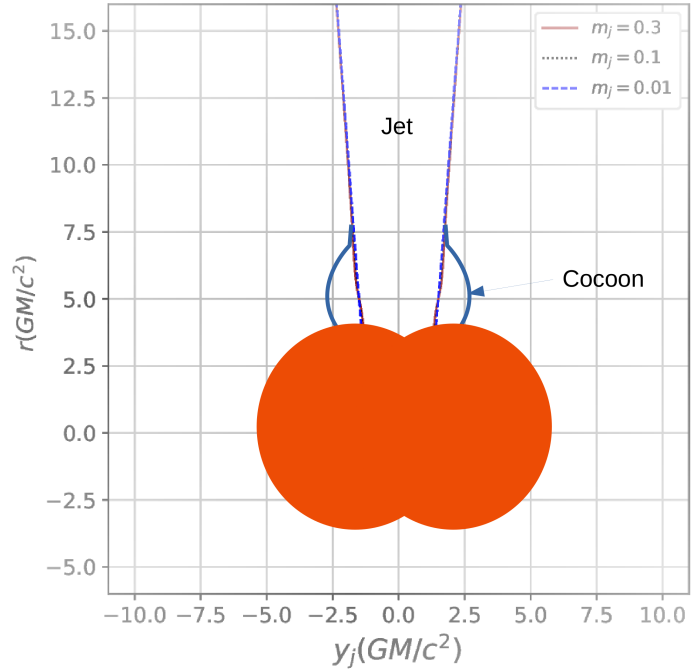}
\caption{The cross-section of the jet above stellar surface for various values of collimation parameters $m=0.01$ (solid red), $0.1$ (dotted black) and $0.3$ (dashed blue). The orange region is the merger where the two compact stars merge. The cocoon shapes the jet geometry and collimates it above the surface of the merger. The lengths are shown in units $GM/c^2$ or 0.5$r_g$.}
\label{lab_jet_geom} 
\end{center}
\end{figure} 

\subsection{Jet cross section}
\label{sec_cross}
A fluid jet composed of baryons and leptons erupts from the stellar surface after the merger and it is shaped by an envelope or cocoon above it. Consider spherical coordinate system $r,\theta, \phi$ whose origin lies at the centre of the merger. The jet propagates vertically along $\theta=0$ axis and the system has azimuthal symmetry (\ie dynamical parameters are independent along $\phi$ coordinate). If the radius of jet stem at radial coordinate $r$ is $y_j=r\sin \theta$, we define the jet geometry as,

\be 
y_j=\frac{mA_1(r-d)}{1+A_2(r-d)^2}+m_{inf}(r-d)+c_j
\label{yj.eq}
\ee
Here the constants are $A_1=0.5,~ ~d=5$, \textit{A}$_2=$1.0. $m_{inf}=0.1/(1+m)$ is the slope of radially expanding jet after crossing the envelope and $c_j=1.5$. $mA_1$ and $A_2$ shape the jet. $d$ constraints the approximate location of the cocoon at $r=d$ (as it is extended and has no precise location). $A_2$ controls the vertical extent up to which the cocoon would affect the jet geometry. We retain constant values of $d$ and $A_2$ in this work, which means that the cork location as well its dimensions are constant. $c_j$ is the mathematical intercept of the jet when it erupts following the merger and is also kept constant. The cross-section of the jet at location $r$ is given as $A=\pi y_j^2$. Here $m$ is a geometric parameter which represents the ability of the cocoon to affect the jet geometry. In other words, $m$ controls the cocoon's strength. This jet geometry is inspired by various studies carried out on an erupting GRB jet (for eg. see \citep{bpb19, ijt21}). For $m=0$, the jet is radial throughout and there is no effect of cork on its geometry. This type of geometric model was used by \citep{vc17} where a jet launched around a black hole is collimated following its interaction with the inner torus of an accretion disc. However, such collimation is more prominent in GRB jet as it pierces the cocoon, hence we choose the parameters accordingly. This considered geometry contains information of a jet that is collimated and then flows radially.

In Figure \ref{lab_jet_geom}, we plot the respective jet geometries corresponding to $m=0.01, 0.1$ and $0.3$ respectively. In the attached cartoon diagram in this figure, the rough jet launching location from the merger is also shown. For higher values of $m$, the jet is shaped after erupting. Hence $m$ controls the effective collimation above the stellar surface due to its interaction with the outer envelope of the star, as well as, the cocoon. The collimation of the jet due to the cocoon leads to the variation in the vertical gradient of the jet cross-section. As we will see in the next section, the parameter that affects the jet dynamics is $A^{-1}dA/dr$. This implies that the slight fractional change in cross-section results in the variation of jet bulk speed.
\begin{figure}[H] 
\begin{center}
\includegraphics[width=10.5 cm]{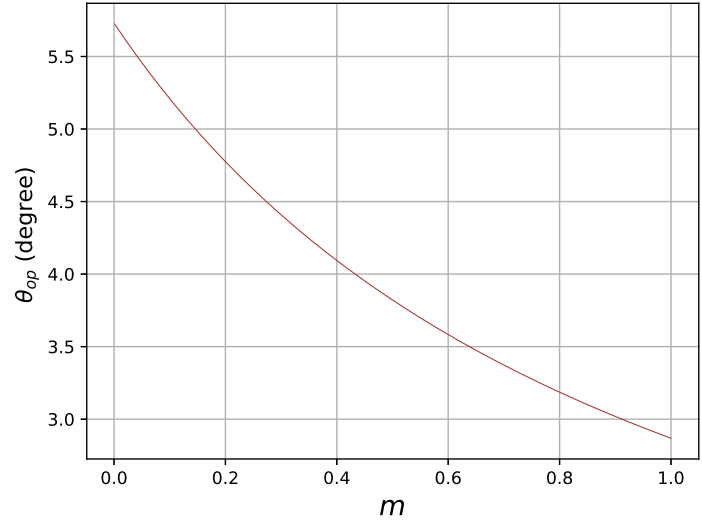}
\caption{Opening angle $\theta_{op}$ as a function of $m$ at $r=100$ (as $r>>d$)}
\label{lab_m_theta_op} 
\end{center}
\end{figure} 
The jet opening angle $\theta_{op}$ is defined at large radius (or $r>>d$ above). To have further physical insight into the jet geometry described above, we write Equation \ref{yj.eq} at large distances in the following form (at a large distance),
$$
y_j=\tan \theta_{op}(r-d)+c_j %{\rm ~~~~[At ~~r>>d]}
$$ 
Here ,
$$
\theta_{op} = \tan^{-1}\left[\frac{A_1m}{1+A_2(r-d)^2}+m_{inf}\right] %{\rm ~~~~[At ~~r>>d]}
$$
In Figure \ref{lab_m_theta_op}, we plot the opening angle of the jet $\theta_{op}$ as a function of $m$ by choosing $r=100$. As expected, a higher value of $m$ leads to stronger collimation of the jet and results in narrow opening angles. This parameterized representation of jet geometry captures the cocoon's ability to collimate the jet. Hence $m$ represents the strength of the cork or cocoon.
This type of collimation is similar to the conventional de Laval nozzle in hydrodynamics \cite{h17}. Here we treat the problem in curved space and the flow is relativistic in nature. In other words, we work out an analysis of the general relativistic de Laval nozzle problem in the context of a GRB jet erupting against a stellar cork ahead of it. 
%\\
\subsection{Relativistic equation of state and its dependence upon flow composition}
\label{sec_eos}
To solve a set of hydrodynamic equations of motion we need an equation of state which is a closure relation between the internal energy density $e$, matter density $\rho$ and fluid pressure $p$. Here we use a relativistic adiabatic equation of state proposed in \citep{rcc06,cr09}. It is an approximated form of the original relativistic equation of state \citep{c57} (Also see \citep{s57}). This equation of state provides a good agreement with exact results (see appendix C of \cite{vkmc15}) and is easy to use in analytic studies. It takes care of multispecies in the fluid flow and has been used to study the effect of flow composition in relativistic fluids \cite{cr09,crj13, vc18a,vc18b, vc19,sc20, ky22,jcr22,pmd22}. It allows us to define the thermodynamic state of the fluid for a given fraction of leptons over baryons. For a relativistic fluid with electron number density $n_{e^{-}}$ the equation of state is given as,
\begin{equation}
e=n_{e^-}m_ec^2f,
\label{eos.eq}
\end{equation}
Here $m_e$ is the mass of the electron. Parameter $f$ is a function of jet composition, as well as, the temperature of the flow.
\begin{equation}
f=(2-\xi)\left[1+\Theta\left(\frac{9\Theta+3}{3\Theta+2}\right)\right]
+\xi\left[\frac{1}{\eta}+\Theta\left(\frac{9\Theta+3/\eta}{3\Theta+2/\eta}
\right)\right].
\label{eos2.eq}
\end{equation}
We defined the non-dimensional temperature as
$\Theta=kT/(m_ec^2)$, with $k$ being the Boltzmann constant. Information of the fluid composition is contained in parameter $\xi = n_{p^{+}}/n_{e^{-}}$ which is a relative proportion of 
number density of protons and electrons. $\xi$ ranges from 0 to 1. $\xi=0$ represents purely leptonic flow composed of electrons and positrons while $\xi=1$ is for electron-proton flow. $\eta = m_{e}/ m_{p^{+}}$ where $m_{p^{+}}$ is the mass of proton.
By definition, the expressions of polytropic index $N$, sound speed $a$ and adiabatic 
index $\Gamma$, as well as, specific enthalpy $h$ are given as,

\begin{equation}
N=\frac{1}{2}\frac{df}{d\Theta} ; ~~
a^2=\frac{\Gamma p}{e+p}=\frac{2 \Gamma \Theta}
{f+2\Theta};~~ \Gamma=1+\frac{1}{N};~~ h =({e+p})/{\rho}=({f+2\Theta})/{\tau}
\label{sound.eq}
\end{equation}
%Further, the specific enthalpy of the fluid is given as,
%$$
%h =({e+p})/{\rho}=({f+2\Theta})/{\tau}
%$$
Here $\tau=(2-\xi+\xi/\eta)$ is a function of jet composition. For an insight into the effect of the relativistic equation of state compared to the nonrelativistic equation of state with an invariant adiabatic index, we direct the reader to compare the results of \citep{vc18a} and \citep{vc18b}, a study carried in the context of X-ray binary jets.

\section{Jet dynamics}
\label{sec_dyn}
\subsection{Dynamical equations of motion of the jet}
The energy momentum tensor of a fluid in bulk motion is given as $T^{\alpha \beta}=(e+p)u^{\alpha}u^{\beta}+pg^{\alpha \beta}$.
%\begin{equation}
%T^{\alpha \beta}=(e+p)u^{\alpha}u^{\beta}+pg^{\alpha \beta}
%\end{equation}
Here $u^{\alpha}$ are the components of four-velocity of the jet fluid and $g^{\alpha \beta}$ are the the metric tensor components. The conservation of energy-momentum enables us to write down the equations of motion by setting the four divergence of $T^{\alpha \beta}$ to zero. \ie
\begin{equation}
T^{\alpha \beta}_{;\beta}=[(e+p)u^{\alpha}u^{\beta}+pg^{\alpha \beta}]_{;\beta} = 0~~~~ % \mbox{and} ~~~~
\label{eq_mo1}
\end{equation}
Projecting Equation \ref{eq_mo1} using the projection operator $(g^{i}_{\alpha}+u^iu_\alpha)$ gives us the momentum balance equation along $i^{th}$ coordinate while projecting it along four velocity $u^{\alpha}$ leads to the energy conservation principle as,
\begin{equation}
(g^{i}_{\alpha}+u^iu_\alpha)T^{\alpha \beta}_{{;\beta}}=0; ~~~ u_{\alpha}T^{\alpha \beta}_{{;\beta}}=0
\label{genmomb.eq}
\end{equation}
\\
For this semi-analytic study, we obtain steady-state equations of motion. We assume an axis-symmetric jet propagating along $\theta=0$ and we need to derive and solve the equations along radial coordinate $r$  as the system has $\phi$ symmetry. As explained in section \ref{sec_assump}, we assume that the jet properties remain invariant along its horizontal extent so we solve the equations along $r$ and reduce the problem to one dimension. The momentum balance equation and the energy conservation equation reduce to,
\begin{equation}
u^r\frac{du^r}{dr}+\frac{1}{r^2}=-\left(1-\frac{2}{r}+u^ru^r\right)
\frac{1}{e+p}\frac{dp}{dr},
\label{eu1con.eq}
%\eqno{(6a)}
\end{equation}
and
\begin{equation}
\frac{de}{dr}-\frac{e+p} {\rho}\frac{d\rho}{dr}=0,
%\eqno{(6b)}
\label{en1con.eq}
\end{equation}
Here $\rho$ is the local fluid density. The mass conservation enables us to write the continuity equation as 
\be 
(\rho u^{\beta})_{; \beta}=0
\label{eq_cont}
\ee
Integration of the continuity equation gives outflow rate along radial coordinate $r$ in spherical coordinates ($r,\theta,\phi$),
\be 
\dot {M}_{\rm {out}}=\rho u^r {A}.
\label{mdotout.eq}
\ee
Here $A$ is the jet cross section. The continuity equation can be written in differential form as,
\be 
\frac{1}{{\rho}}\frac{d{\rho}}{dr}+\frac{1}{{A}}\frac{d{A}}{dr}
+\frac{1}{u^r}\frac{du^r}{dr}=0
%\eqno{(6c)}
\label{con1con.eq}
\ee
Pressure $p$ is given as
\be
p=\frac{2\Theta \rho}{\tau}=\frac{2\Theta \dot {M}_{\rm {out}}}{\tau u^r A}
\label{pressure.eq}
\ee
using these relations, the equations of motion convert to,

\begin{equation}
\frac{dv}
{dr}=\frac{\left[a^2\left\{\frac{1}{r(r-2)}+\frac{1}{A}
\frac{dA}{dr}\right\}-\frac{1}{r(r-2)}\right]}{\gamma^2v\left(1-\frac{a^2}{v^2}\right)}
\label{dvdr.eq}
\end{equation}
and
\begin{equation}
\frac{d{\Theta}}{dr}=-\frac{{\Theta}}{N}\left[ \frac{{\gamma}
^2}{v}\left(\frac{dv}{dr}\right)+\frac{1}{r(r-2)}
+\frac{1}{A}\frac{dA}{dr}\right]
\label{dthdr.eq}
\end{equation}
$v$ is the three velocity of the jet. It is defined as
\be 
v^2=-u_iu^i/u_tu^t=-u_ru^r/u_tu^t %{~~\rm or~~} 
\ee
Here $u^r=\sqrt{g^{rr}}{\gamma}v$ and $u_t=-{\gamma}\sqrt{(1-2/r)}$. $\gamma^2=-u_tu^t$ is the Lorentz factor and $g^{rr}=1-2/r$.
We need to integrate equations \ref{dvdr.eq} and \ref{dthdr.eq} simultaneously to obtain the dynamical evolution of the jet parameters $v$ and $\Theta$ along radial coordinate $r$. All the other parameters are then derived using the relations defined above in the text.
\subsection{Constants of motion}
Integrating momentum balance equation gives the first constant of motion, \ie
\be
A (e+p)u^ru_t=-{\dot E}={\rm constant}
\label{energflux.eq}
\ee
We obtain the specific energy by dividing equation (\ref{energflux.eq}) to (\ref{mdotout.eq})
\ie
\begin{equation}
E=\frac{{\dot E}}{{\dot M}_{\rm out}}=-hu_t. %+\int \frac{(2-\xi)R_d}{g^{rr}r^2\gamma^2}dr
\label{enr.eq}
\end{equation}
It also enables us to define the total kinetic power of the jet as,
\be
L_j={\dot E}={\dot M}_{\rm out}E
\label{ljet.eq}
\ee
Integrating energy conservation equation (\ref{en1con.eq}) one obtains an adiabatic relation similar to
$p\propto \rho^{\Gamma}$ for a constant $\Gamma$ with,
\be
\rho={\cal C}\mbox{exp}(k_3) \Theta^{3/2}(3\Theta+2)^{k_1}
(3\Theta+2/\eta)^{k_2},
\label{rho.eq}
\ee
Here the defined parameters are, $k_1=3(2-\xi)/4$, $k_2=3\xi/4$, $k_3=(f-\tau)/(2\Theta)$ while ${\cal C}$ is the entropy constant. We obtain the entropy outflow rate by substituting $\rho$ into
equation (\ref{mdotout.eq}),
\begin{equation}
{\dot {\cal M}}=\frac{{\dot M}_{\rm out}}{{\rm geom. const.}
{\cal C}}=\mbox{exp}(k_3) \Theta^{3/2}(3\Theta+2)
^{k_1}
(3\Theta+2/\eta)^{k_2}u^rA
\label{entacc.eq}
\end{equation}
The two constants of motion are defined by equations (\ref{entacc.eq}) and (\ref{enr.eq}). ${\dot {\cal M}}$ is discontinuous at the shock.

\section{Method of obtaining solutions : sonic point analysis and shock conditions}
\label{sec_method}
Once erupted from the surface of the merger, the matter is hot and sub-relativistic which implies that the matter has to be subsonic ($v<a$). As it progresses, it speeds up due to thermal pressure and cools down subsequently. Escaping to infinity, the jet is supersonic ($v>a$). At some distance $r=r_s$, the jet has to pass through a sonic point where the local sound speed equals the bulk jet speed. Flows associated with such solutions are called transonic flows. The entropy of these solutions is maximum among the whole family of solutions. Following the second law of thermodynamics, the jet naturally chooses the transonic trajectory having maximum entropy. It is a necessary condition for the escape of a subsonic flow to become relativistic.  We inject the jet with sub-relativistic and subsonic speeds and solve equations \ref{dvdr.eq} and\ref{dthdr.eq} simultaneously using Runge Kutta 4th order method for the numerical solutions. Stability of the coupled equations and the test of convergence of the numerical solutions require that the constants of motion $E$ and $\dot {\cal M}$ should remain conserved across the jet extent and it should always be checked for each solution. Alternatively, one can study the dynamical evolution of the system directly using equations \ref{enr.eq} and \ref{entacc.eq} to arrive at identical results.

Sonic point analysis is important in studying the mathematical nature of the solutions, as well as, it helps in revealing their physical insight. The location of a sonic point is obtained when the denominator of equation \ref{dvdr.eq} equals zero. The sonic point condition is thus,
\be 
a_s^2=\left[1+r_s(r_s-2)\left(\frac{1}{{A}}\frac{d{A}}{dr}\right)_s\right]^{-1}\
\label{sonic.eq}
\ee
At $r=r_s$, the slope $dv/dr_{(r=r_s)}=0/0$ is undetermined and we need to use L'hopital's rule for its estimation \citep{s63}.
%\ie
%\be 
%\frac{dv}{dr}=\frac{N/dr}{dD/dr}
%\ee
For a given sonic point, the energy parameter, as well as, entropy is determined. With condition $v_s=a_s$,
the sonic point serves as a mathematical boundary from which the transonic solutions can be alternatively computed by integrating the equations of motion outward ($dr\rightarrow +ve$), as well as, inward ($dr\rightarrow -ve$).

We will see that the collimation of the jet by the cocoon leads to the formation of multiple sonic points for a range of energy parameter $E$. In the current case, we will see that a jet can have up to three sonic points, which makes it possible to go through a shock transition. The shock conditions across the shock are given by relativistic Rankine-Hugoniot equations where the following flow quantities remain constant across the flow \cite{t48}.
\begin{equation}
  [{\rho}u^r]=[(e+p)u^tu^r]=[(e+p)u^ru^r+pg^{rr}]=0,
  \label{sk1.eq}
\end{equation}
%\begin{equation}
%   [T^{tr}]=[(e+p)u^tu^r]=0,
%   \label{sk2.eq}
%\end{equation}
%and
%\begin{equation}
%[T^{rr}]=[(e+p)u^ru^r+pg^{rr}]=0
%\label{sk3.eq}
%\end{equation}
The quantities in square brackets represent their differences across the shock. \ie $[A]=A_2-A_1$, where subscript 1(2) denotes the respective quantity in pre-shock (post-shock) region. These conditions ensure the invariance of mass, momentum and energy fluxes across the shock. In case multiple sonic points are present for a given value of $E$, the solutions from outer and inner sonic points can have a shock transition between them if the values of fluxes in Equation \ref{sk1.eq} coincide at shock location $r=r_{sh}$. The shock is always located between the outer and inner sonic points. For any given value of $E$, followed by multiple sonic points, we check for the solutions from outer and inner sonic points and test for consistency from the shock conditions defined above to look for the location of shock transition between the two solutions. It may be noted that the entropy of the flow does have a discontinuous jump at the shock and is not necessarily conserved in such a case. The fluid is compressed at the shock and the associated compression ratio is defined as
\be 
{\cal R} = \frac{\rho_2}{\rho_1}
\ee
Higher value of $\cal R$ represents stronger shock transition. From the continuity equation (Equation \ref{mdotout.eq}), we can write it as
\be 
{\cal R} = \frac{\gamma_1 v_1}{\gamma_2 v_2}>1
\ee
\begin{figure}[H]
\begin{center}
\includegraphics[width=13.5 cm]{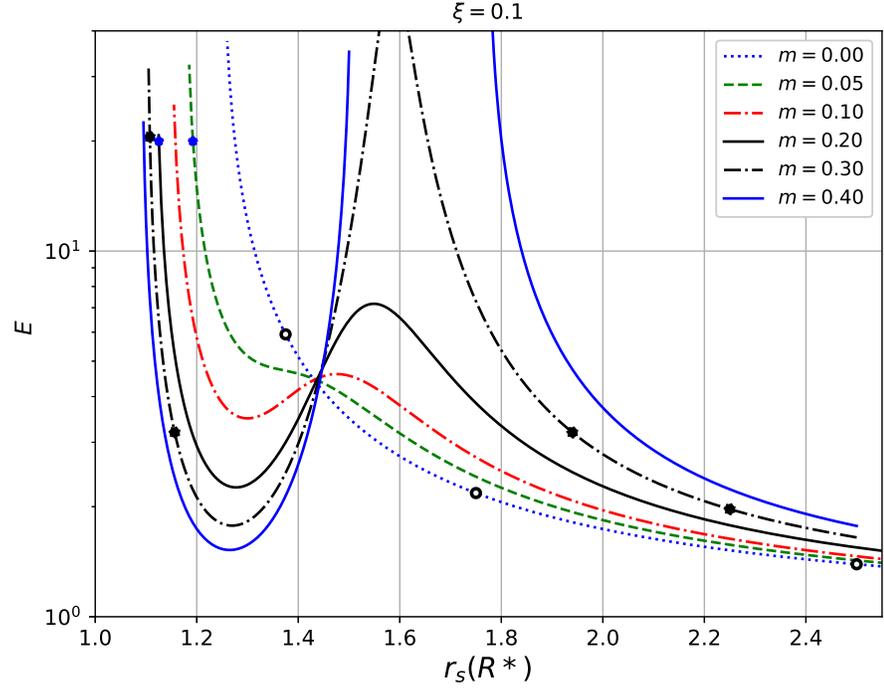}
\caption{$E-r_s$ parameter space for various values of $m=0.0, 0.05, 0.1, 0.2, 0.3$ and $0.4$ keeping $\xi=0.1$. Deviation from monotonous trajectory for $m=0$ marks effect of cocoon's collimation on the jet dynamics. For $m\geq 0.1$, in certain range of $E$, we obtain three sonic points for single value of $E$, it marks possibility of shock transition in the flow. Points with black open circles and filled blue stars show the location of sonic points corresponding to solutions in Figure \ref{lab_vel_m0} top left panel and bottom right panel respectively. Solutions corresponding to sonic points with filled black circles on case $m=0.3$ are shown in Figure \ref{lab_vel_m0.3_xi_0.1}. }
\label{lab_sonic_1} 
\end{center}
\end{figure}
\section{Results}
\label{sec_results}
For a given energy parameter $E$, the total energy of the jet is fixed as we work with the adiabatic equation of state. We inject the jet at the surface of the star assigning it a value of $E$ which corresponds to sub relativistic flow speeds and high sound speed. As the jets are thermally driven in this study, their dynamic evolution for a given energy parameter is likely to be affected by their composition ($\xi$), as well as, the cocoon's strength which is controlled by the parameter $m$ in the jet cross-section profile. %The family of transonic solutions is studied through $E-r_s$ parameter space. 
\begin{figure}[H]
\begin{center}
\includegraphics[width=6.5 cm]{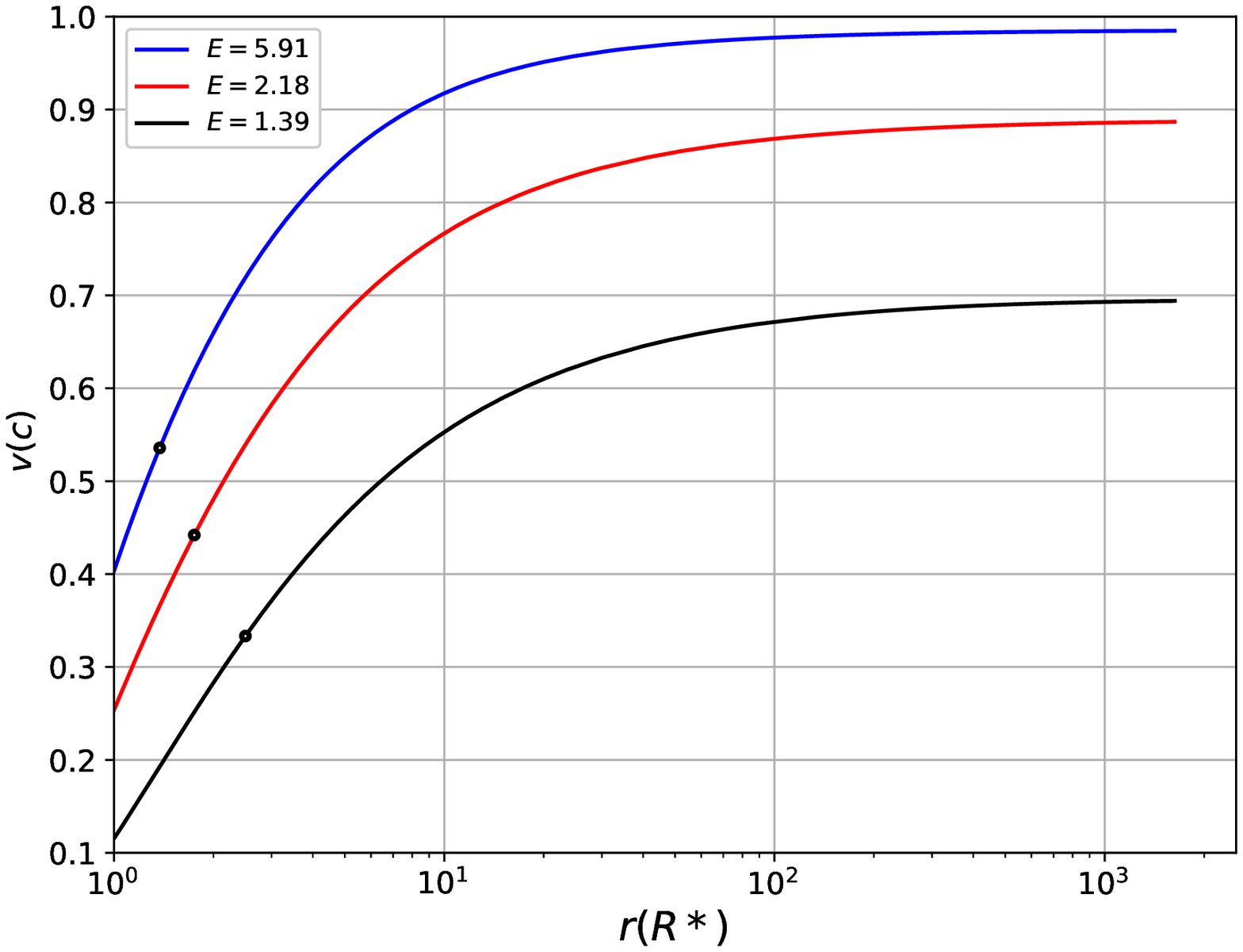}
\includegraphics[width=6.5 cm]{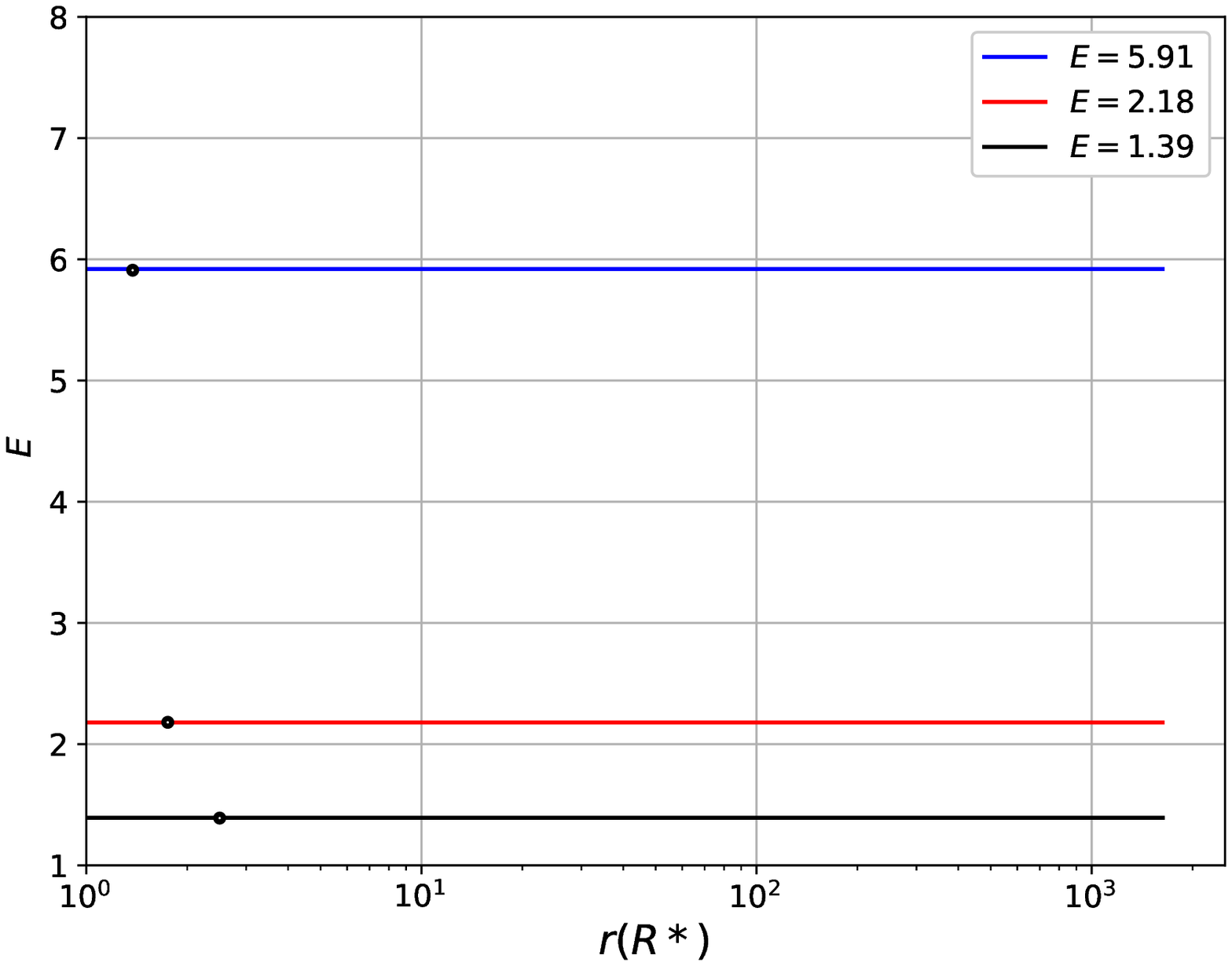}
\includegraphics[width=6.5 cm]{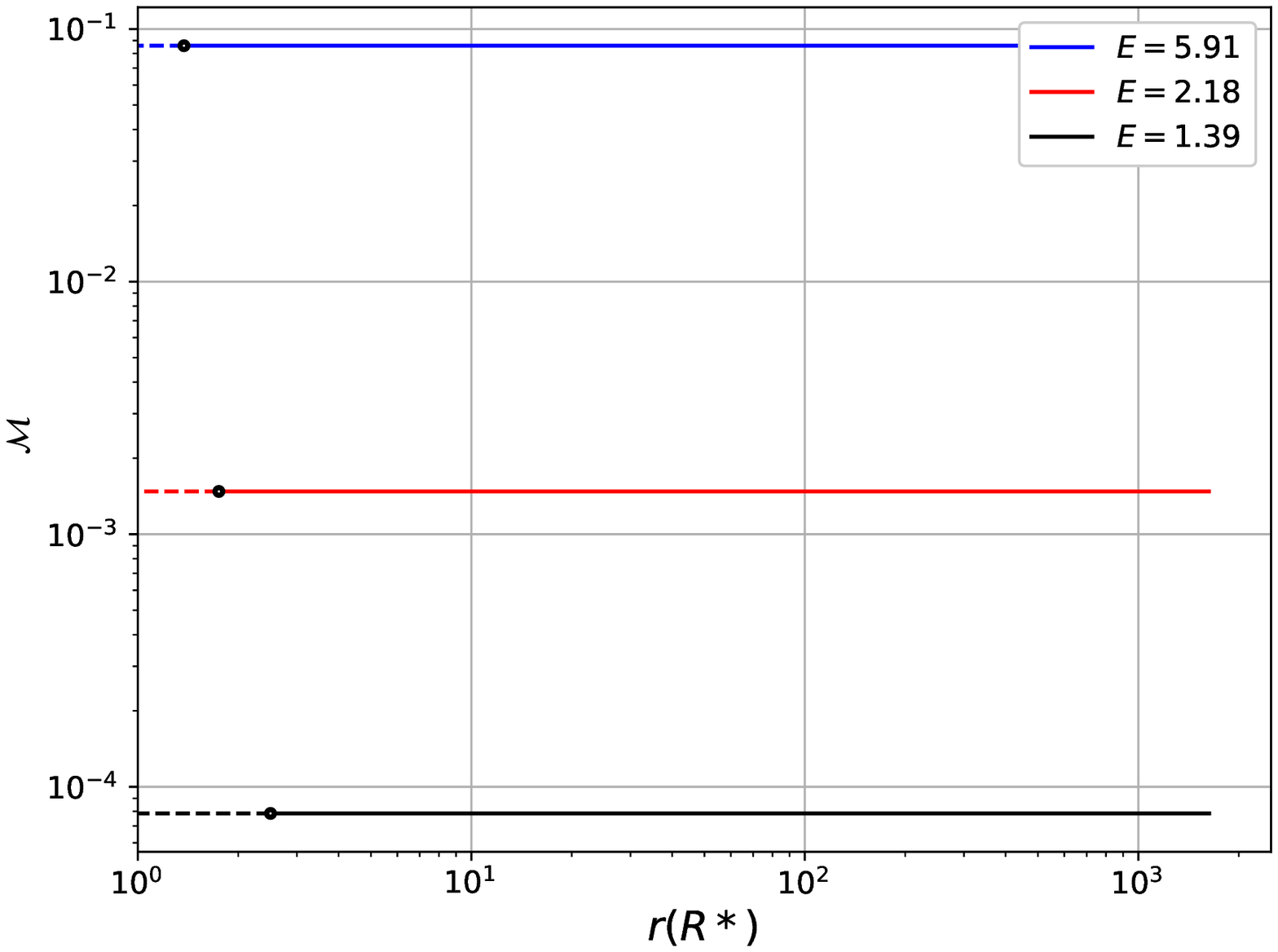}
\includegraphics[width=6.5 cm]{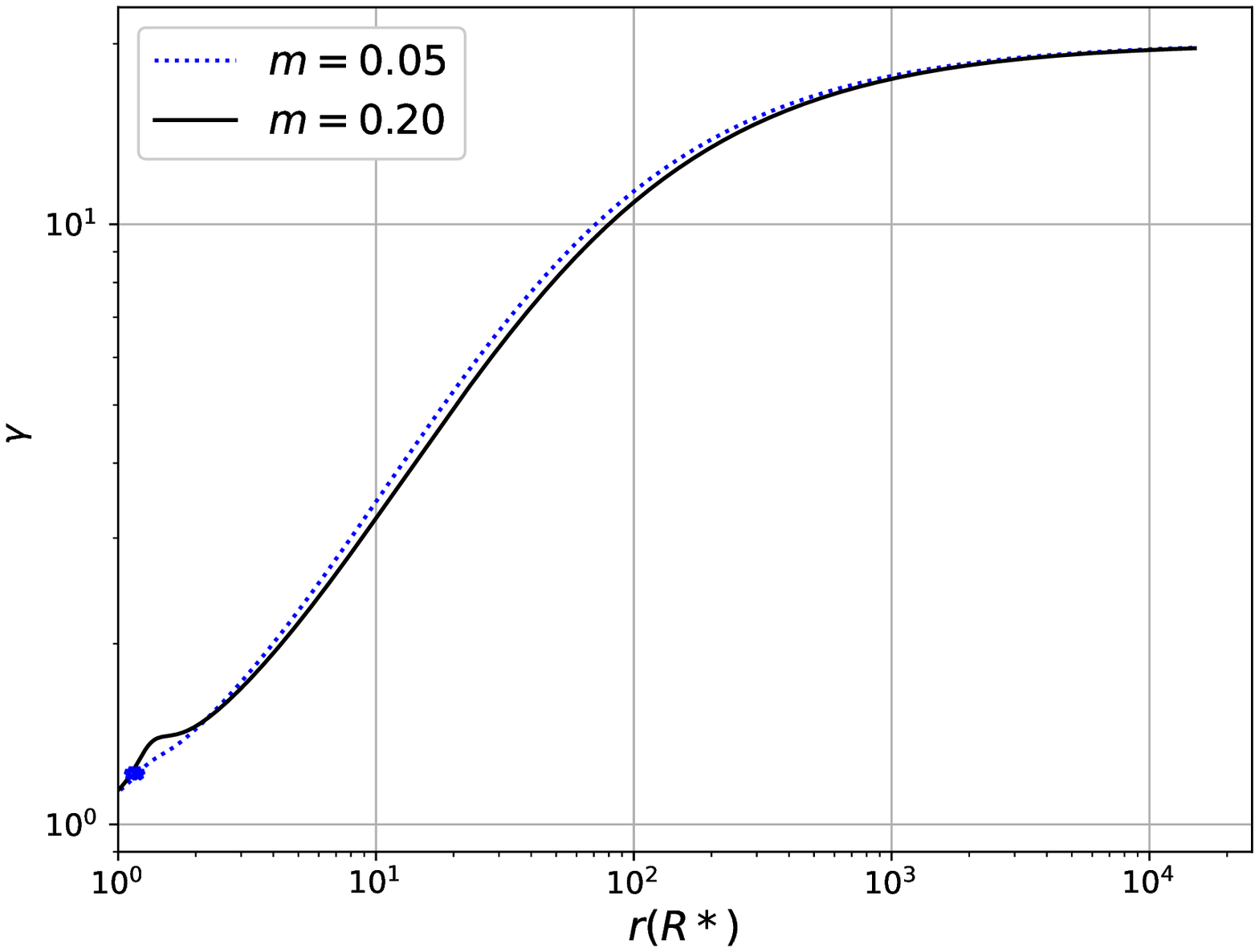}
\caption{Top left : Velocity profiles associated with Figure \ref{lab_sonic_1} for case $m=0$ choosing different values of $E=5.91$ (blue), 2.18(red), 1.39 (black); Establishing the invariance of the constants of motion $E$ (top right) and ${\dot {\cal M}}$ (bottom left) along radial coordinate $r$ for solutions in Figure \ref{lab_sonic_1}. Bottom right : Lorentz factor profiles for the choices of smaller values of $m=0.05$ and $0.2$ choosing $E=20$. Points with black open circles and blue filled stars denote the locations of sonic points as in Figure \ref{lab_sonic_1}. For all panels above, $\xi=0.1$. }
\label{lab_vel_m0}
\end{center}
\end{figure}  
\subsection{Dependence of flow solutions on the cocoon's strength}
To study the effect of the jet cross-section on its dynamics we set to analyze a family of sonic points for different values of the energy parameter keeping a constant composition. In Figure \ref{lab_sonic_1} we plot the energy parameter $E$ as a function of sonic point $r_s$ for various choices of $m$. For these results, we consider a fixed composition assigned by $\xi=0.1$. 

For $m=0$, the jet is radial in nature and $E$ monotonously decreases with $r_s$. It means that a single sonic point corresponds to each value of $E$ and there is no effect of the cocoon on the jet's dynamics. It produces monotonous jet velocity ($v$) profiles, plotted in Figure \ref{lab_vel_m0} (top left panel) for various values of energy parameter $E=5.91,2.18$ and $1.39$. $E\rightarrow1$ represents a jet with $\gamma\rightarrow1$ at infinity. Hence $E$ is  realizable for values $E\geq1$. Higher energy content shows more relativistic jets. Along with ${\dot M}_{\rm out}$ [Equation \ref{ljet.eq}] $E$ constraints the total kinetic luminosity of the jet. Both the constants of motion $E$ (top right) and $\dot{\cal M}$ (bottom left) remain invariant across the spatial extent of the jet in all the solutions. Black open circles in these velocity profiles mark the location of the corresponding sonic point. These points are also shown in Figure \ref{lab_sonic_1}. The higher energy parameter brings the sonic point closer to the surface of the ejecta which is seen in Figure \ref{lab_sonic_1} (blue dotted curve).
These trajectories are similar to conventional Bondi solutions in relativistic regime.  For higher values of $m$ (\ie $m\geq 0.1$), a single value of $E$ corresponds to multiple sonic points which leads to the possibility of shock transition. For a more detailed account of the nature of multiple sonic points and the formation of shocks, see \citep{kh76,ht83,ftr85}.
In the bottom right panel of Figure \ref{lab_vel_m0}, we plot the Lorentz factor profile for high energy jet with $E=20$ and a weak cocoon (\ie small values of $m=0.05,$ and $0.2$). The jet becomes transonic at points shown by filled blue circles.
The cocoon slightly affects and collimates the jet after its ejection. However, the terminal value of $\gamma$ at infinity is unaffected by the cocoon's presence. This fact directly follows from the expression of $E$ and our consideration of the adiabatic equation of state where no energy dissipation is involved in the jet's propagation throughout its spatial extent.

\begin{figure}[H]
\begin{center}
\includegraphics[width=12.5 cm]{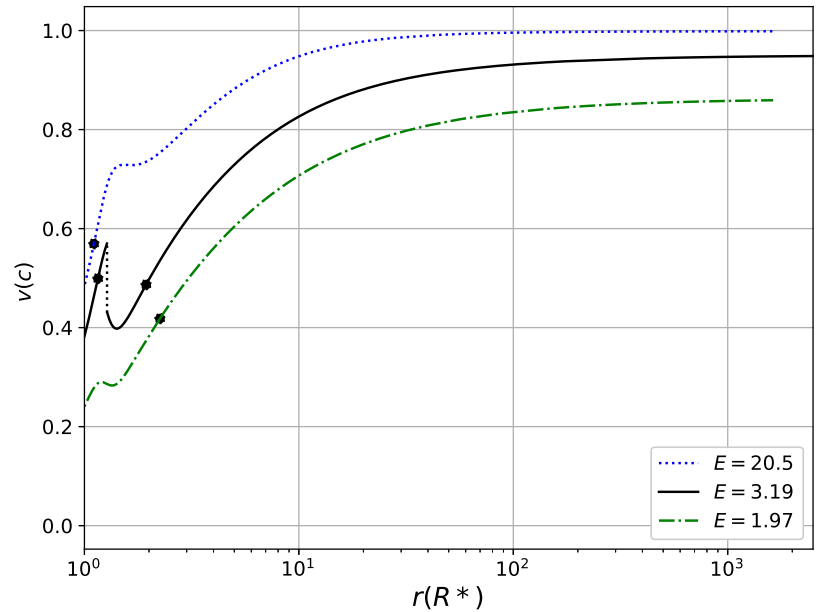}
\caption{Velocity profiles for $m=0.3$ with different values of $E=20.5$ (dotted blue), 3.19 (solid black) and 1.97 (dashed-dotted green). The vertical dotted line marks the shock transition at location $r=1.275R^*$ and filled circles denote the sonic points through which the physical solutions pass.}
\label{lab_vel_m0.3_xi_0.1}
\end{center}
\end{figure}  
Next, we choose a stronger cocoon by setting $m=0.3$ and keeping $\xi=0.1$ same as before. The corresponding velocity profiles are plotted in Figure \ref{lab_vel_m0.3_xi_0.1} for three different values of jet energies $E=20.5$( Dotted blue), 3.19 (solid black) and $1.97$ (green dashed-dotted). Physical solutions pass through sonic points shown by black filled circles (Also marked in Figure \ref{lab_sonic_1}). From Figure \ref{lab_sonic_1}, in $E-r_s$ parameter space for $m=0.3$, we observe that multiple sonic points are present for a single value of $E$ in large sonic point space ($r_s<2$). Thus, it leads to the possibility of a shock transition. From the velocity profiles, we see that a jet with high energy ($E=$20.5, blue dotted) doesn't care much about the cocoon's presence and is mildly decelerated by cocoon interaction before further accelerated due to thermal pressure to achieve relativistic speeds. Similarly, a jet with a very low energy parameter (E=1.97, red dashed-dotted) is also not able to form shocks and has a smooth solution up to infinity. However, at intermediate energies ($E\sim 3.19$, black solid) the jet goes through recollimation shock transition at $r_{sh}=1.275 R^*$. The reason why low energy jets are able to escape without being much affected by the cocoon is that they do not have a sufficient momentum to interact with the cocoon to go through a discontinuous transition. It should be noted that the jets with even low energies ($E\rightarrow 1$) are chocked by the cocoon or generate breeze solutions. We do not consider such solutions in this study and restrict ourselves to the cases where the jet has minimum energy to convert into transonic outflows with a successful escape.

\begin{figure}[H]
\begin{center}
\includegraphics[width=6.5 cm]{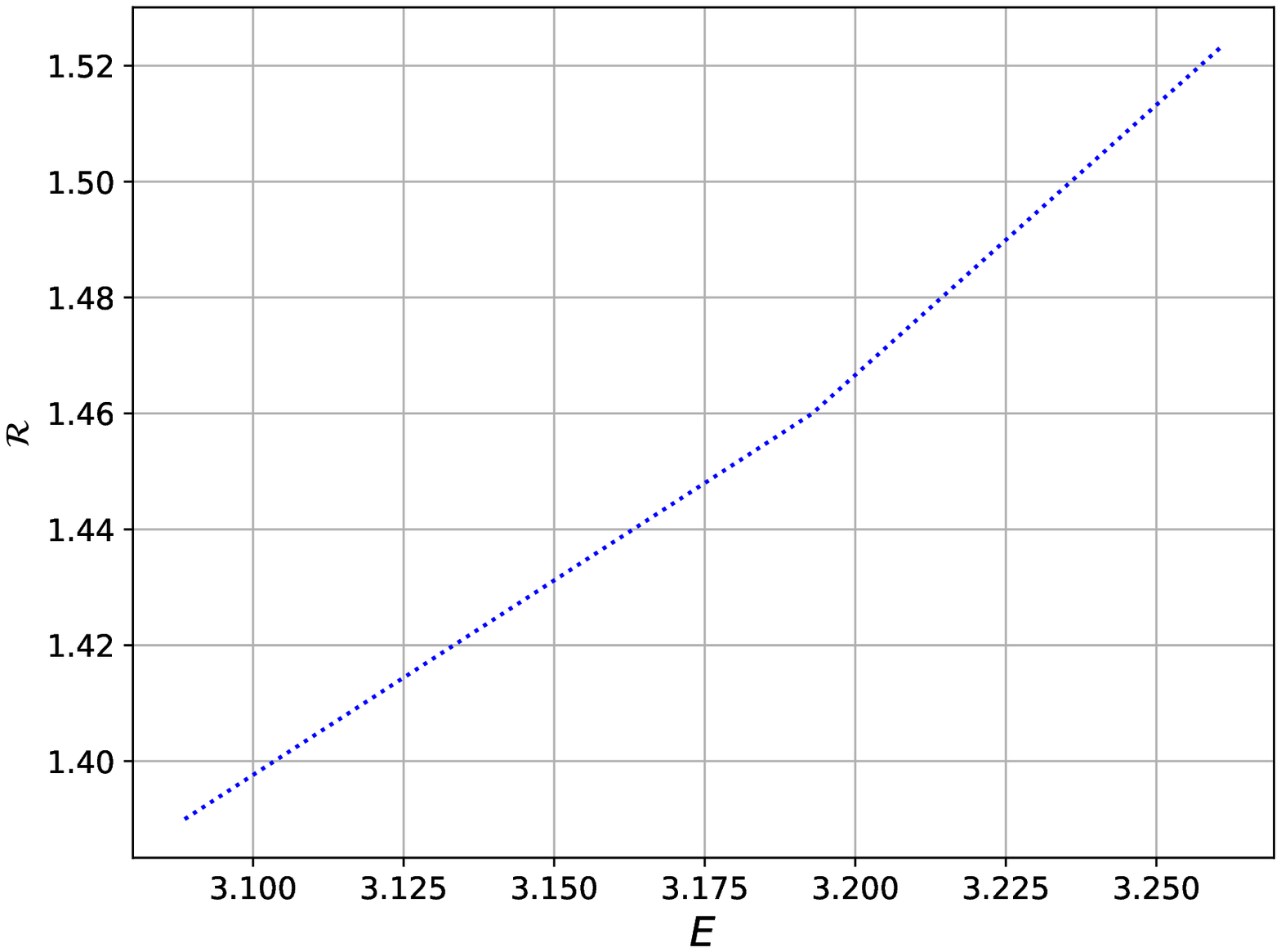}
\includegraphics[width=6.5 cm]{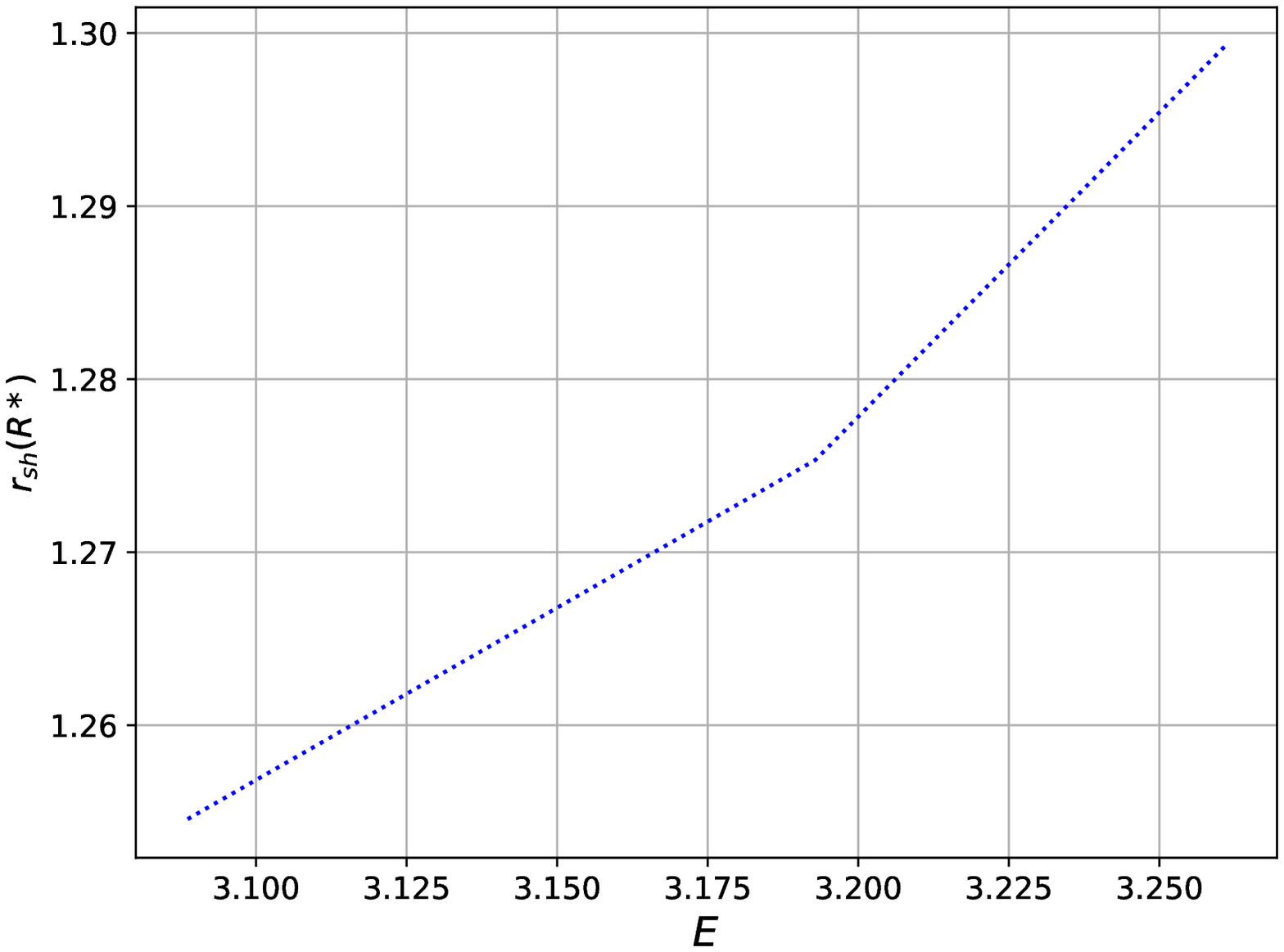}
\includegraphics[width=6.5 cm]{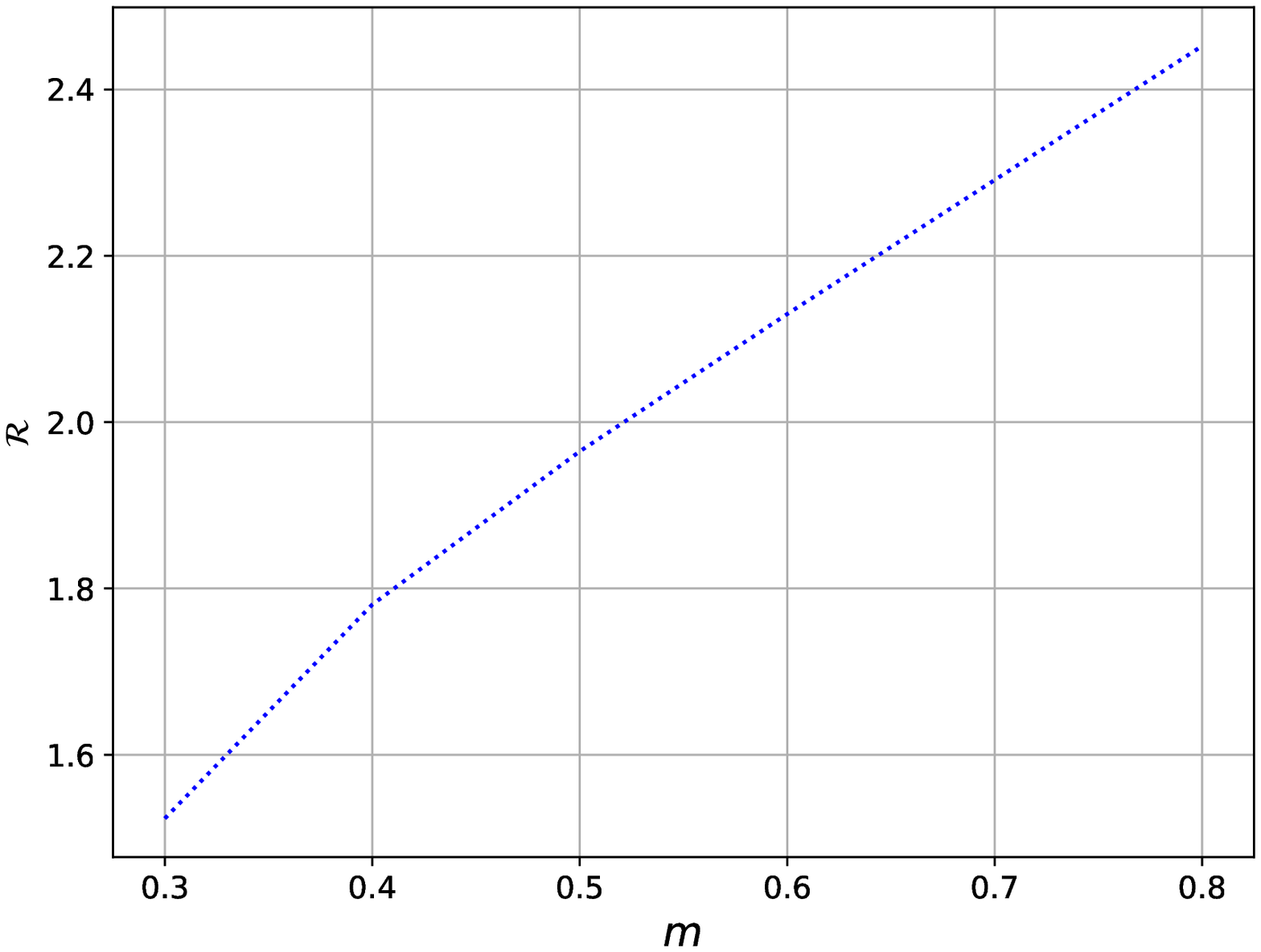}
\includegraphics[width=6.5 cm]{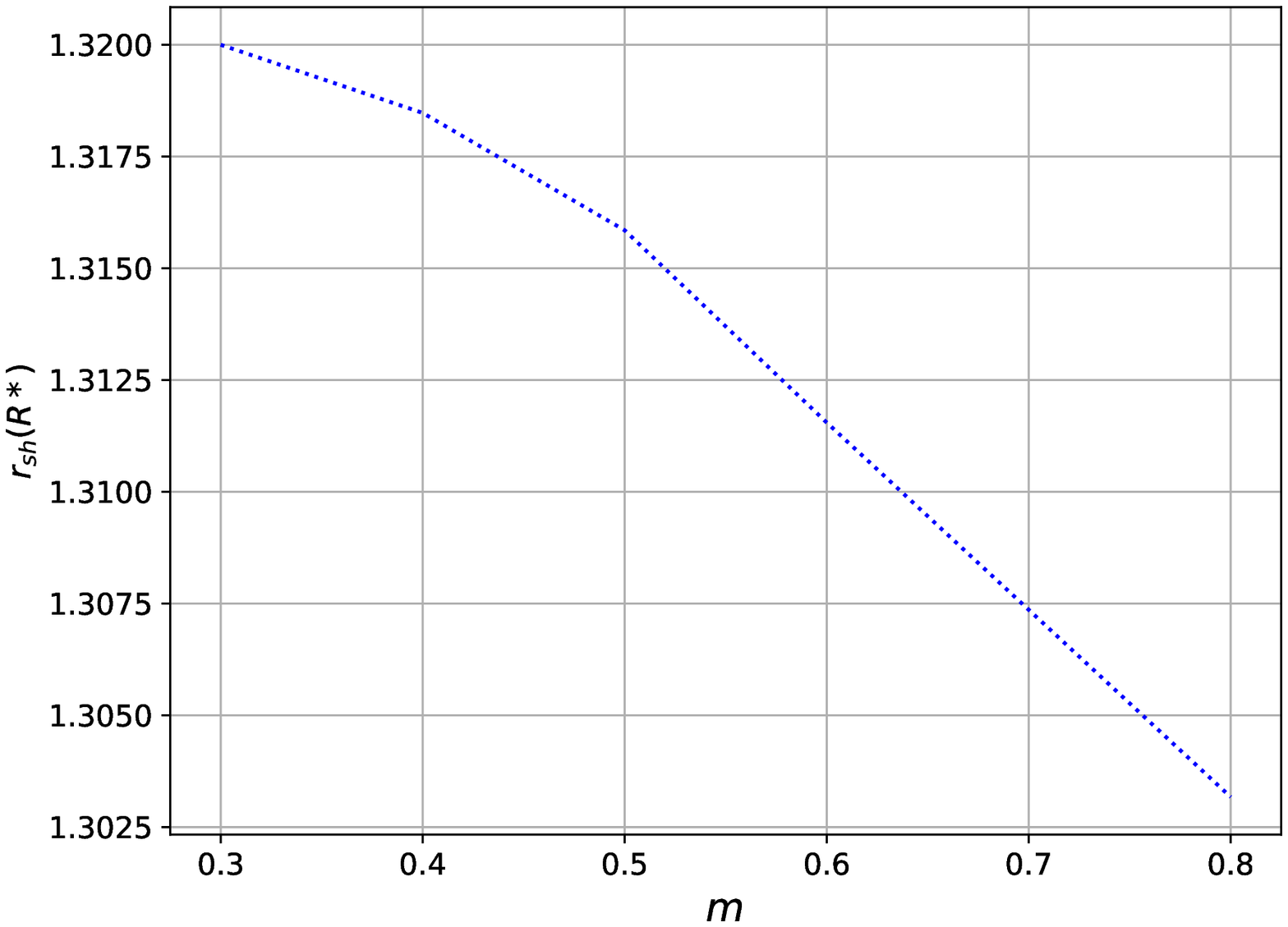}
\caption{Variation of compression ratio $\cal R$ (top left) and shock location $r_{sh}$ (top right) with $E$ for $m=0.3$. Similarly $\cal R$ (bottom left) and $r_{sh}$ (bottom right) as functions of $m$ keeping $E=3.26$. For all these plots, $\xi=0.1$.}
\label{lab_R_vs_E}
\end{center}
\end{figure}

The jet harbours shocks within the energy range $E=3.09$ to $3.2$. In Figure \ref{lab_R_vs_E} we plot the compression ratio $\cal R$ as a function of $E$ (top left panel) and show that the shock in the high energy jets is stronger. In top right panel the variation of shock location $r_{sh}$ is showed with $E$. The high value of $E$ pushes the shock away from the base and the spatial shock region appears between $r=1.25R^*$ and $1.3R^*$ above the stellar surface. These solutions are for a fixed strength of cocoon (or a constant $m=0.3$). In bottom left and bottom right panels of Figure \ref{lab_R_vs_E}, we plot $\cal R$ and $r_{sh}$ as functions of $m$ keeping a jet with fixed energy parameter $E=3.26$. As an obvious result, we obtain a stronger shock for higher values of $m$ or a stronger cocoon. Subsequently, a stronger cocoon produces shocks closer to the jet base (\ie $r_{sh}$ decreases with increasing $m$).

\begin{figure}[H]
\begin{center}
\includegraphics[width=6.7 cm]{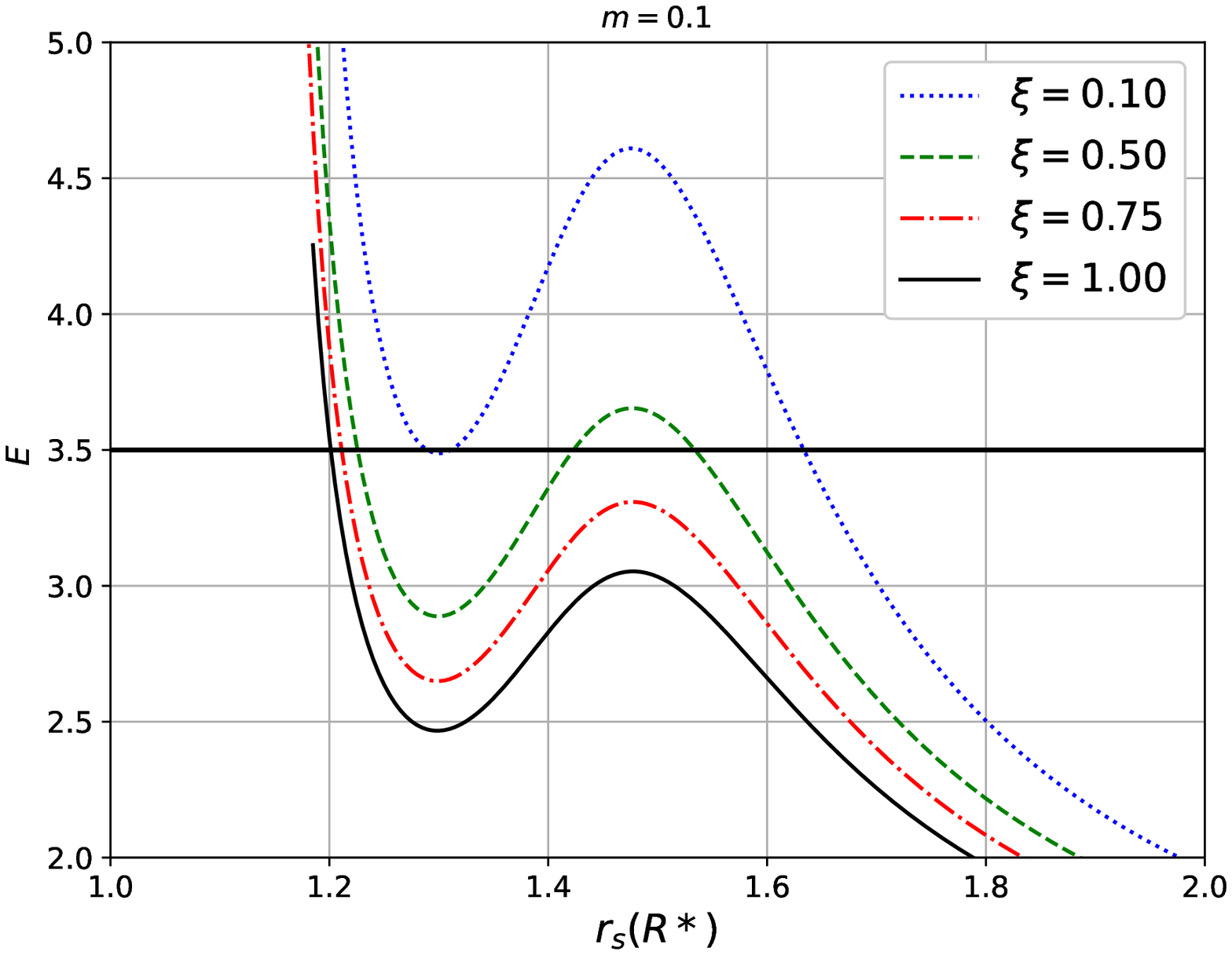}
\includegraphics[width=6.7 cm]{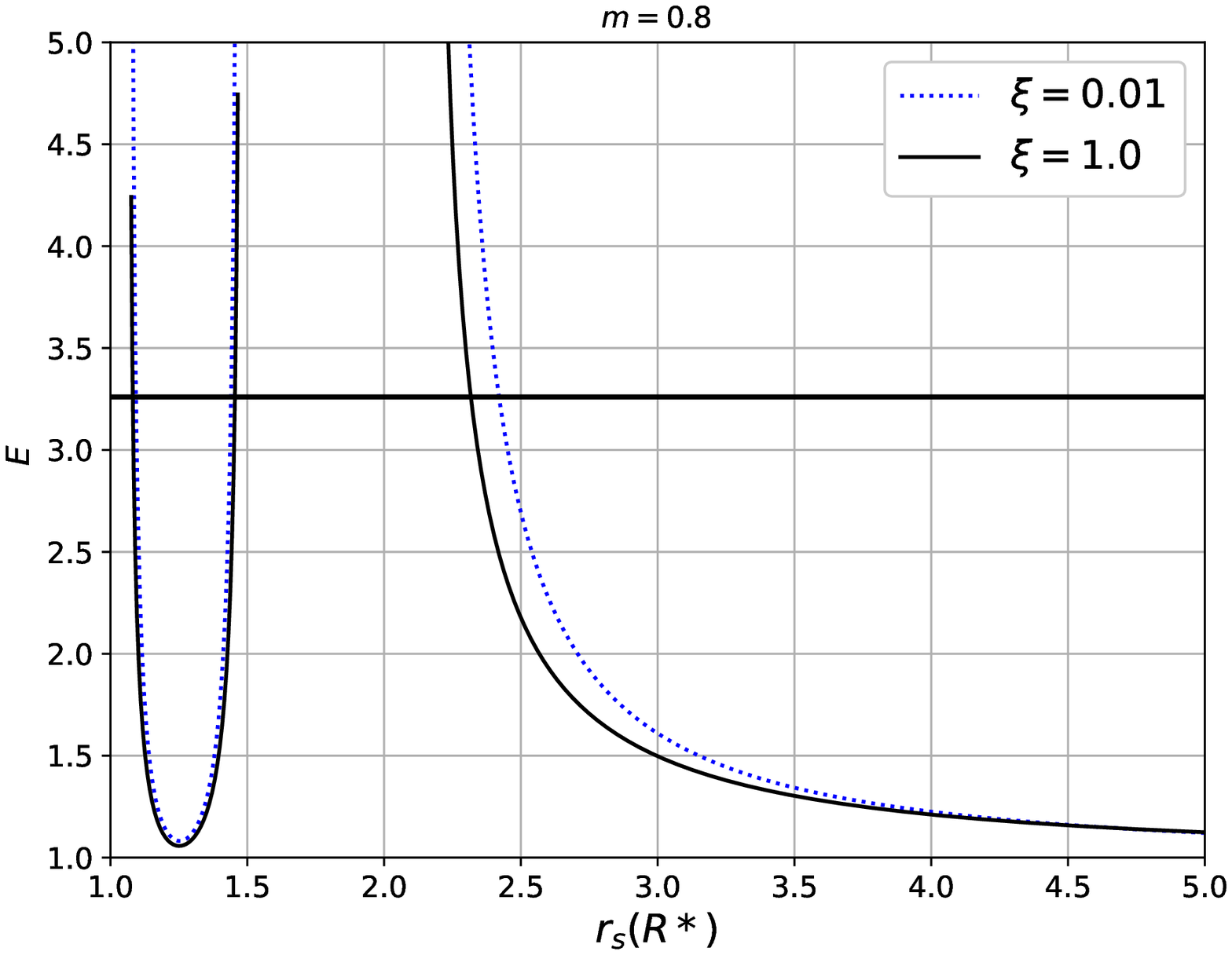}
\caption{$E-r_s$ parameter space for various choices of jet composition. Left panel : $\xi=0.1$ (dotted blue) 0.5 (dashed green), $0.75$ (dashed dotted red) and 1.0 (solid black) for a moderate cocoon strength with $m=0.1$. Velocity profiles corresponding to $E=3.5$ (horizontal black solid line) are plotted in Figure \ref{lab_vel_m0.1_E_3.5_xi_vary}. 
Right panel: $\xi=0.01$ (dotted blue) and 1.0 (solid black). Solutions of $E=3.26$ (horizontal black solid line) are shown in Figure \ref{lab_vel_m0.8_xi_1.0}.}
\label{lab_sonic_2_xi_vary}
\end{center}
\end{figure}  
\subsection{Effect of fluid composition on jet dynamics}
To investigate the effect of flow composition on the jet dynamics, we choose $m=0.1$, that corresponds to a weak cocoon and plot $E-r_s$ parameter space in Figure \ref{lab_sonic_2_xi_vary} (left panel) for different choices of the jet composition $\xi=0.1$ (dotted blue), 0.5 (dashed green), 0.75 (dashed-dotted red) and 1.0 (solid black). For whole range of $\xi$, the jet harbours multiple sonic points within distance $\sim1.2R^*-1.7R^*$. However, the occurrence of multiple sonic points is more prominent in jets with low $\xi$. It can be understood as the low value of $\xi$ corresponds to a relatively high fraction of leptons over baryons. Or the jet has a lower inertia. The cocoon more effectively collimates a less dense jet. In the right panel, we consider strong cocoon with $m=0.8$ for $\xi=0.01$ (solid blue) and $\xi=1.0$ (dashed black). As the cocoon is stronger, it effectively collimates the jet irrespective of its composition and harbours multiple sonic points in most of the parameter space ($r_s<5R^*$).
\begin{figure}[H]
\begin{center}
\includegraphics[width=12.5 cm]{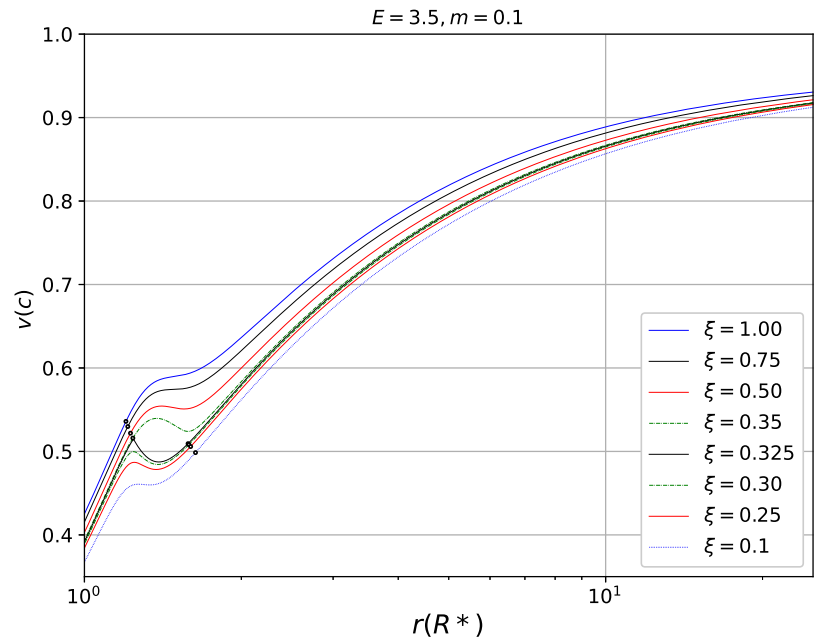}
\caption{Jet velocity profiles for collimation parameter $m=0.1$ and $E=3.5$ for different choices of $\xi$ in range $0.1-1$. Curves with decreasing vertical position represent a decreasing value of $\xi$. No shock solutions are found as the collimation is weak.}
\label{lab_vel_m0.1_E_3.5_xi_vary}
\end{center}
\end{figure}  
In Figure \ref{lab_vel_m0.1_E_3.5_xi_vary}, we plot the velocity profiles associated with weaker cocoon with $m=0.1$ (Figure \ref{lab_sonic_2_xi_vary} left panel) and for a fixed value of $E(=3.5)$ marked by horizontal black line in the parameter space for $\xi$ in range 0.1-1.0. Solutions from top to bottom are in the order of decreasing value of $\xi$. Interestingly, heavy jets ($\xi=1.0$) and lepton dominated jets ($\xi=0.1$) are less affected by the cocoon compared to intermediate values of $\xi$. Jets with high and low inertia are able to pierce through the cork more easily than the jets with intermediate inertia. However, the cocoon is weak and it is not capable of inducing shocks in the flow for all the parameters.
 
 \begin{figure}[H]
\begin{center}
\includegraphics[width=6.7 cm]{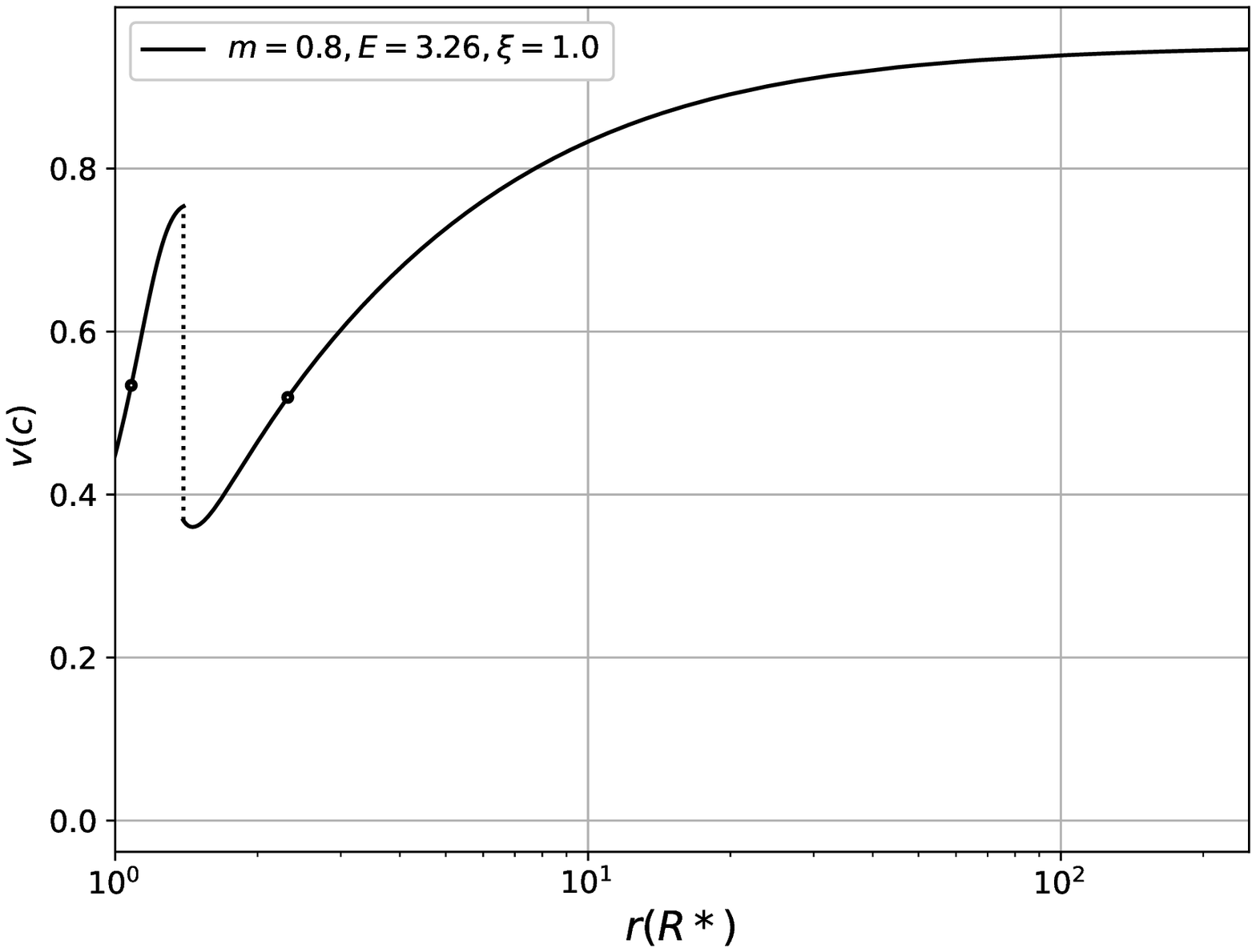}
\includegraphics[width=6.7 cm]{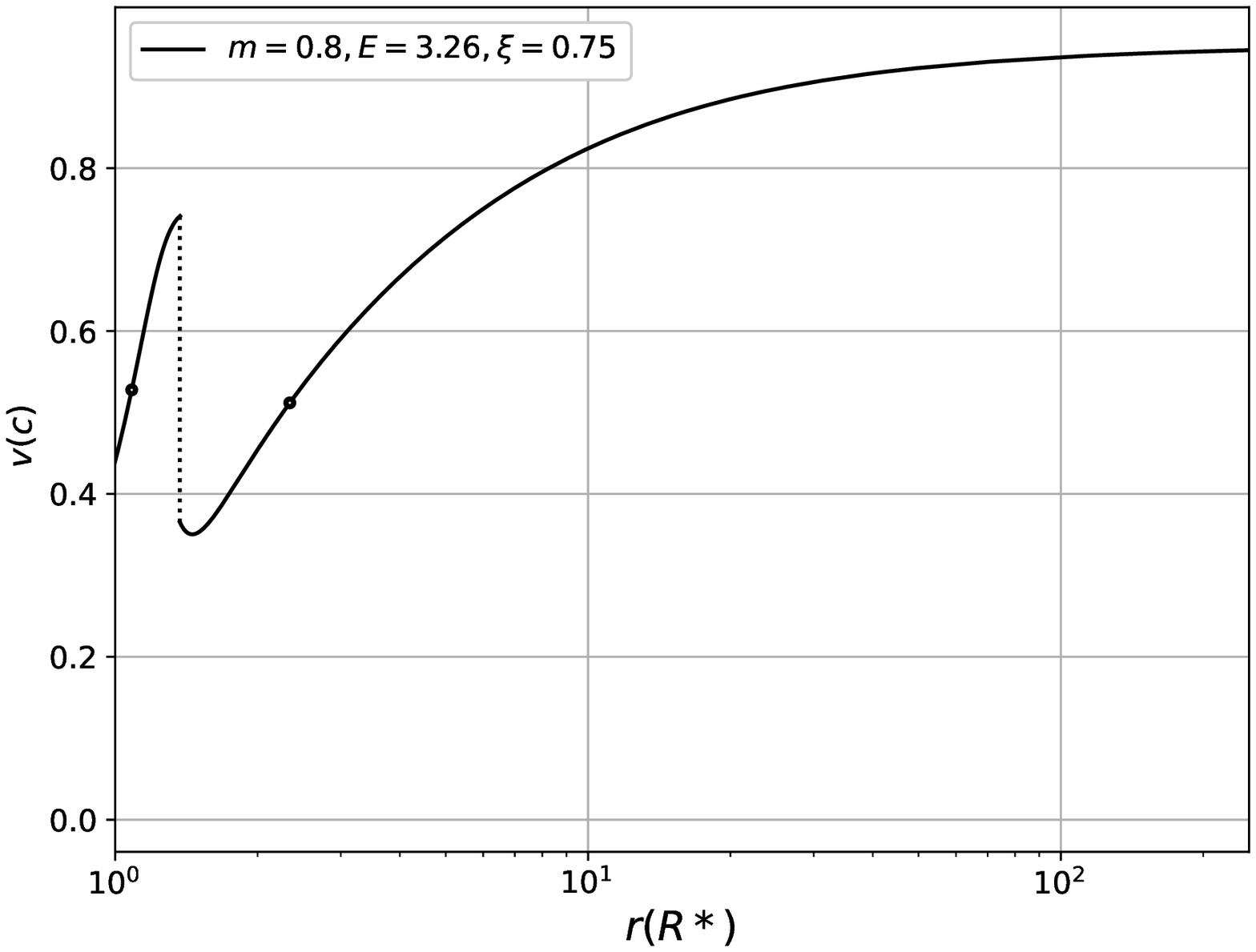}
\includegraphics[width=6.7 cm]{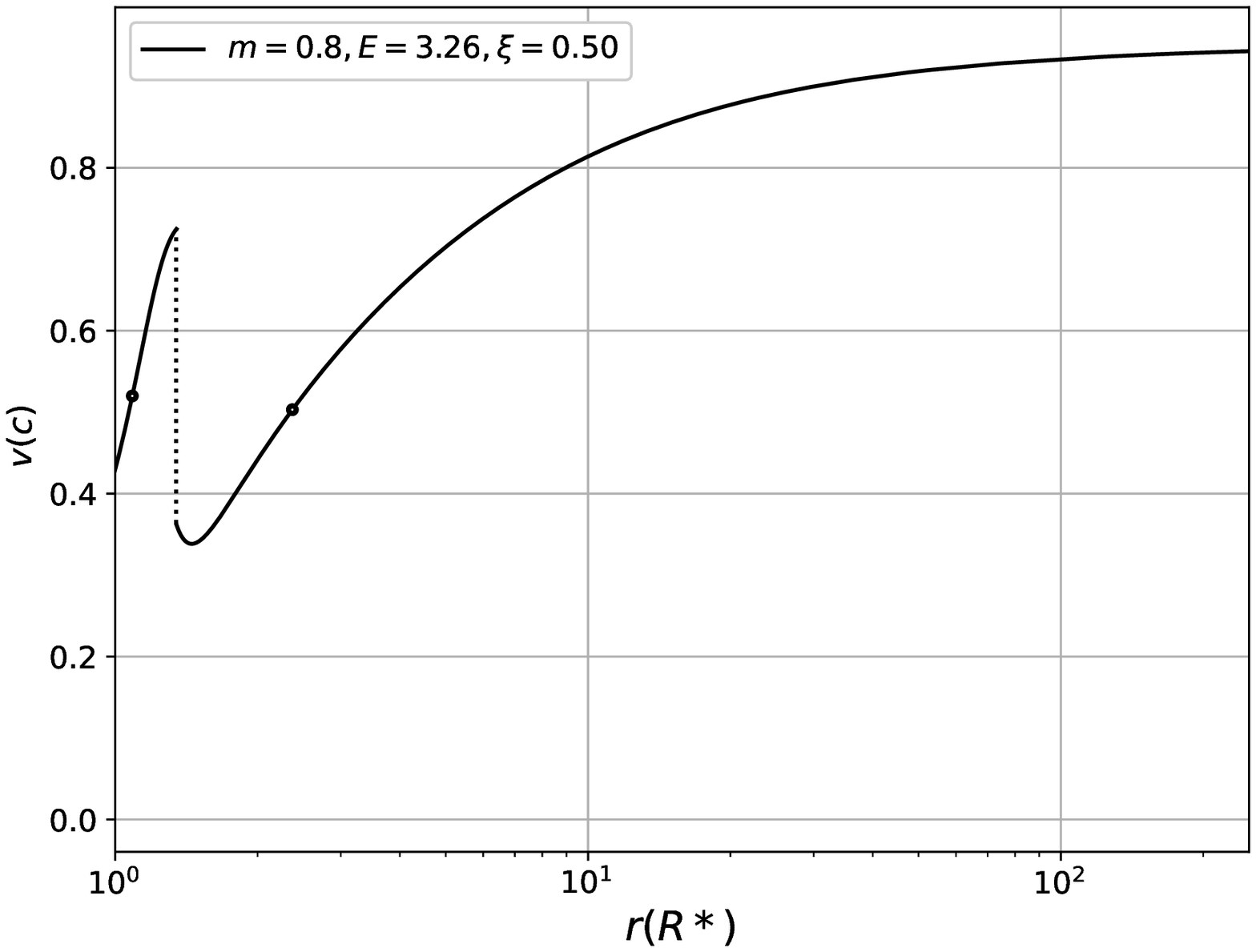}
\includegraphics[width=6.7 cm]{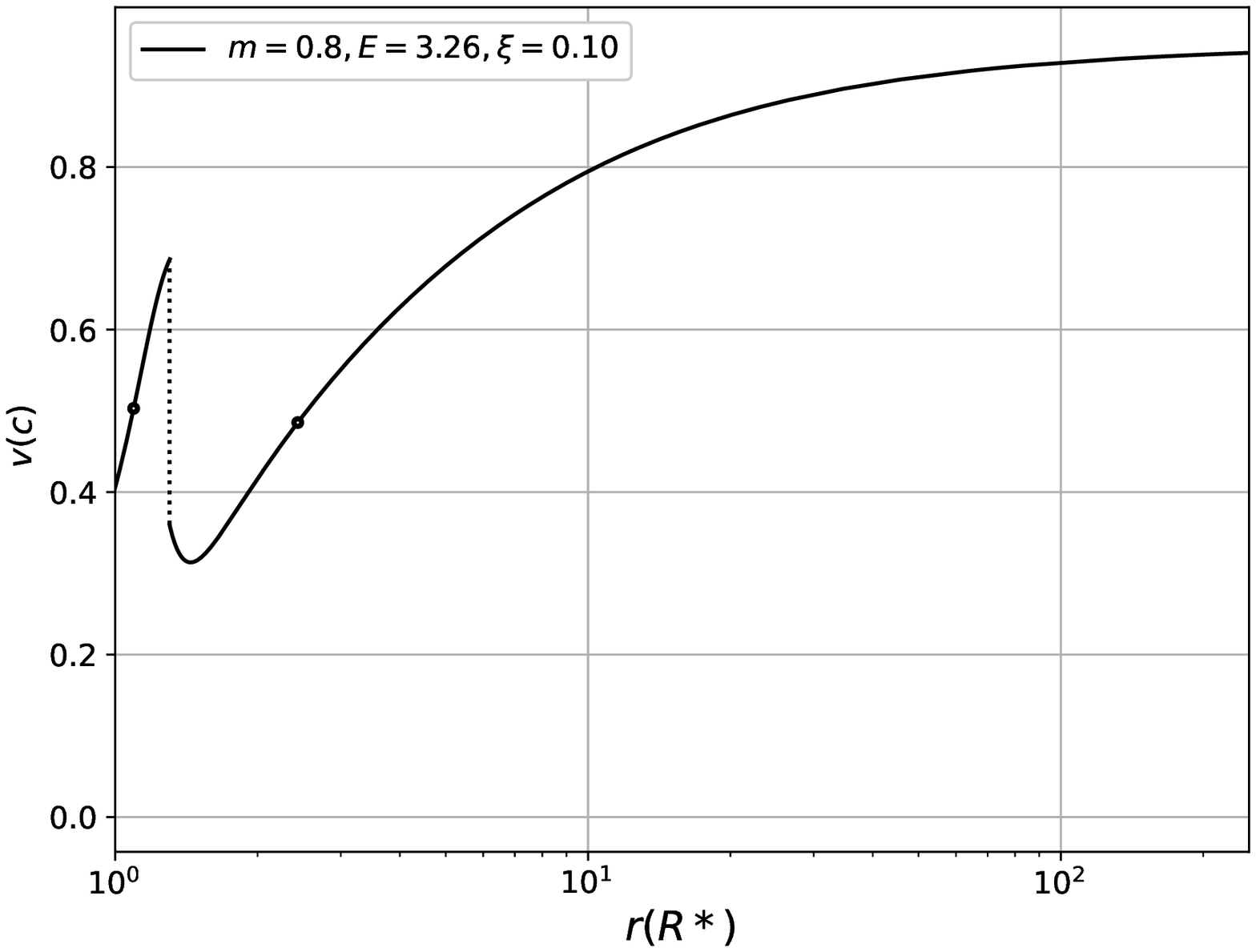}
\includegraphics[width=10.5 cm]{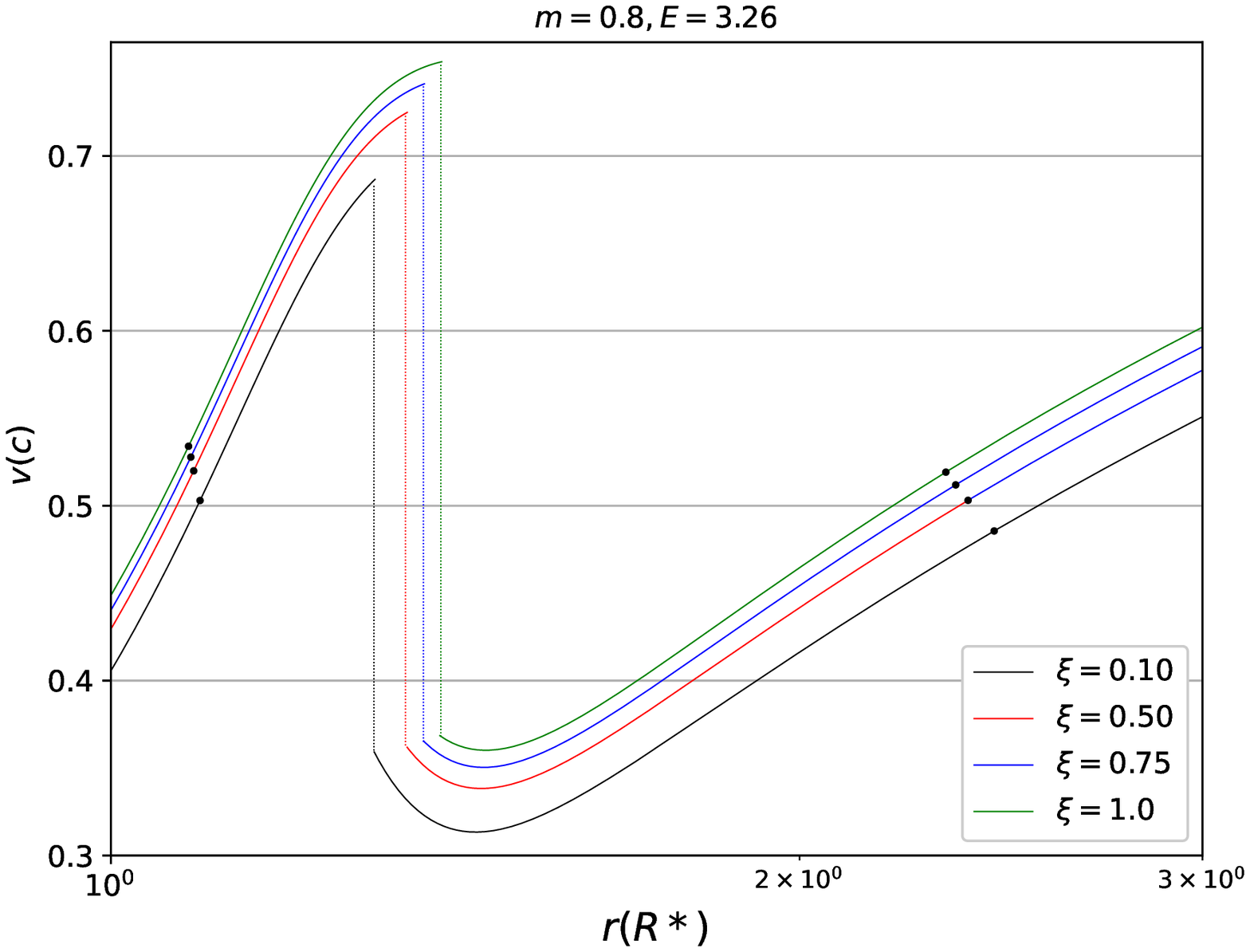}
\caption{Jet velocity $v$ for different choices of $\xi=1.0$ (top left), 0.75 (top right), 0.50 (middle left), and 0.10 (middle right)  with fixed values of $m=0.8$ and $E=3.26$. The whole range of $\xi$ produces shock in the jet as the collimation is strong. In the bottom panel, we overplot these profiles and zoom the region near the surface of the star to show that the shock gets stronger and is pushed apart as the baryon fraction in the jet increases.
}
\label{lab_vel_m0.8_xi_1.0}
\end{center}
\end{figure}  
\begin{figure}[H]
\begin{center}
\includegraphics[width=7 cm]{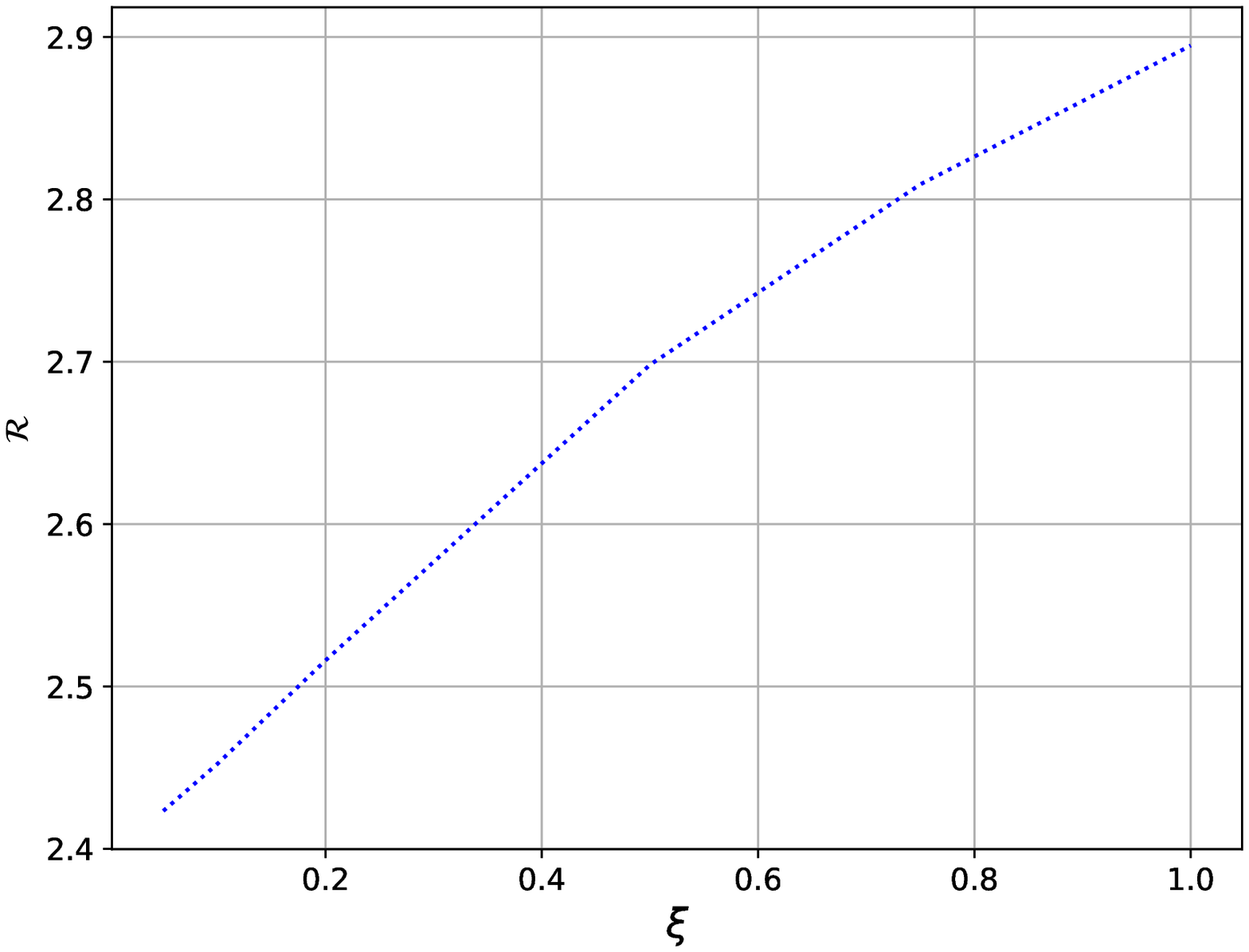}
\includegraphics[width=7 cm]{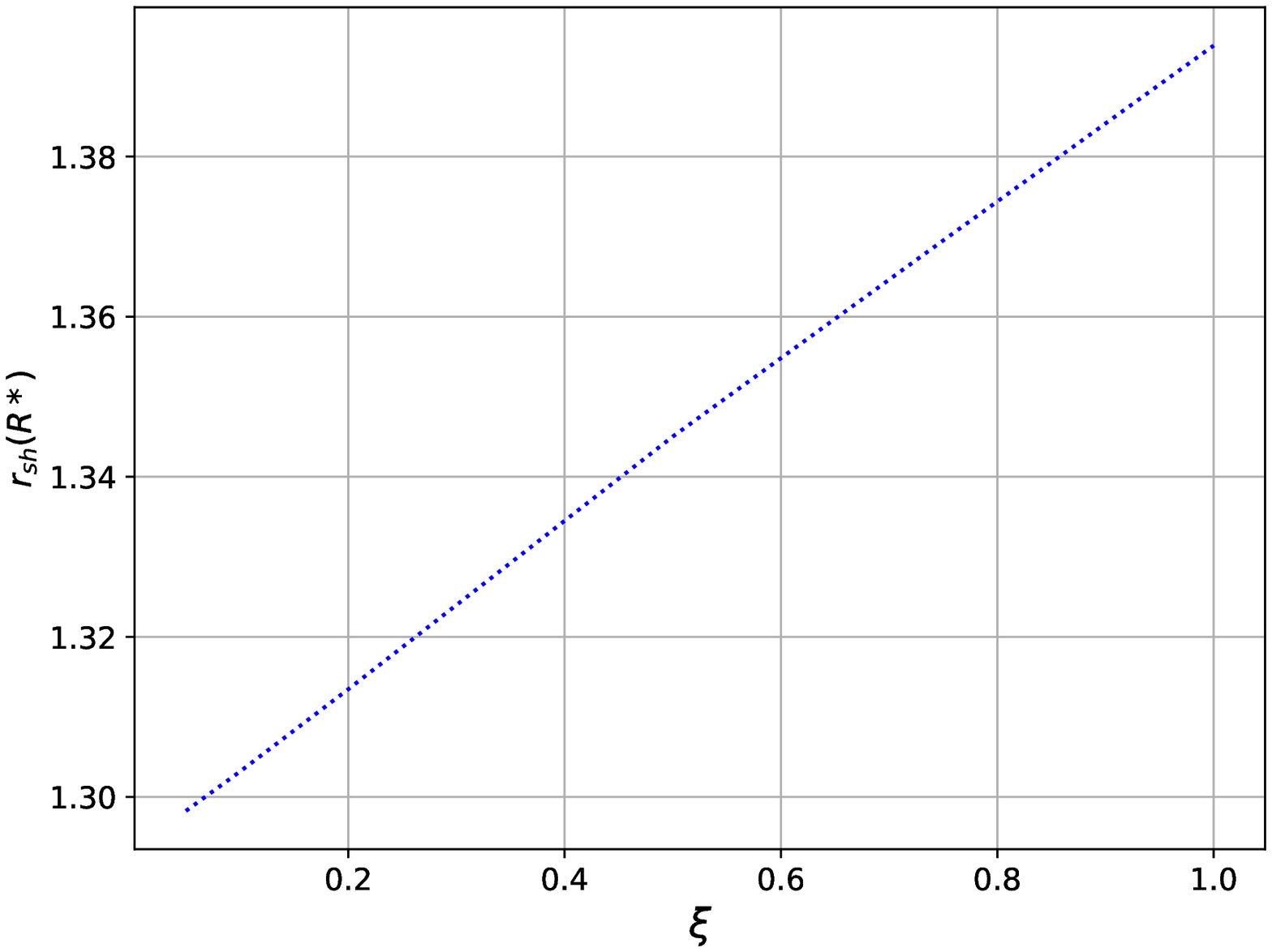}
\caption{Variation of compression ratio $\cal R$ (bottom left) and shock location $r_{sh}$ (bottom right) with composition parameter $\xi$. whole range of $\xi$ produces shock in the jet as the collimation is strong.}
\label{lab_R_vs_xi}
\end{center}
\end{figure} 

 In Figure \ref{lab_vel_m0.8_xi_1.0}, we plot the velocity profiles for a jet piercing a strong cocoon with $m=0.8$. Energy of the jet is kept fixed at $E=3.26$ (the horizontal black line in Figure \ref{lab_sonic_2_xi_vary} right panel) and four cases of $\xi=1.0$ (top left), $\xi=0.75$ (top right), $\xi=0.50$ (middle left) and $\xi=0.1$ (middle right) are chosen. As the cocoon interacts with the jet strongly, in all the cases, the jet goes through a shock transition between the outer and inner sonic points marked by black open circles. In the bottom panel, we over-plot all these solutions and zoom in on the location around shock generation. The shock is pushed away from the jet base for an increasing value of $\xi$ and subsequently, the shock transition is stronger which is seen in the height of the vertical shock transition dashed curves. This conclusion is quantitatively visible in Figure \ref{lab_R_vs_xi} where we plot the variation of compression ratio $\cal R$ (left) and shock location $r_{sh}$ (right) with $\xi$. Jet with a higher baryon fraction produces stronger shock while the shock subsequently forms at larger distances.

\subsection{Jets with supersonic injection}
\begin{figure}[H]
\begin{center}
\includegraphics[width=12.5 cm]{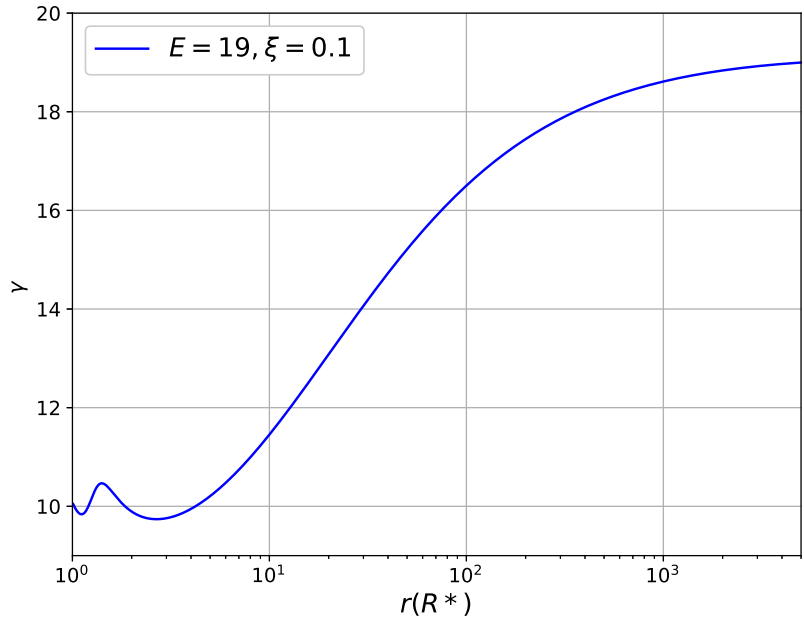}
\caption{Jet Lorentz factor profile $\gamma$ for jets with supersonic injection with $\gamma=10$ at $R=R_*$. Here $m=0.8$, $\xi=0.1$ and $E=19$.}
\label{lab_vel_super}
\end{center}
\end{figure}

So far we discussed transonic jets that start with subsonic speeds and become supersonic at the sonic point. There is another possibility for the jets that are injected at the stellar surface with supersonic speeds or relativistic Lorentz factors \citep{mi13}. Such jets travel at supersonic speeds throughout their propagation. We consider one such case in Figure \ref{lab_vel_super}. A jet with an initial Lorentz factor $10$ is considered with an energy parameter $E=19$. It faces a cork with collimation parameter $m=0.8$. The flow composition is kept at $\xi=0.1$. We observe that the jet shows deceleration due to the cocoon collimation. After escaping, it reaches Lorentz factors $\approx 20$. Such jets are physically realizable if there are accelerating agents that drive the jet up to supersonic speeds at the time of eruption from the surface of the star.

\section{Discussion and conclusions}
\label{sec_conclusions}
In this work, we studied a gamma-ray burst produced by the merger of a compact binary star and giving rise to a relativistic jet that breaks out of the surface of the merger. The jet faces a cocoon or stellar envelope above the surface and pierces through it to escape to infinity. In the process, the jet dynamics is affected and it is effectively collimated by the cocoon. We describe the cocoon's effectiveness by an empirical jet geometry in this model (Equation \ref{yj.eq}) where the cocoon's effectiveness is controlled by the collimation parameter $m$. This geometry sets the defined cross-section that the jet follows during its evolution. We solve the hydrodynamic equations of motion with the help of adiabatic relativistic equation of state which is sensitive to the jet composition. The fluid composition is controlled by parameter $\xi$ and it assigns the presence of lepton fraction over baryon concentration in the fluid.  This is an exploratory analysis where we describe the jet dynamics within all possible ranges of jet energy parameter $E$, cocoon strength $m$ and the lepton fraction in its matter composition.

In this hydrodynamic study of thermally driven relativistic GRB jets, we reconfirmed that a sufficiently strong cocoon is able to produce recollimation shocks in the jet stem. Additionally, we explored the dependence of the shock properties on dynamical parameters of the system. We can draw the following conclusions from this analysis.
\begin{itemize}
\item The mechanical interaction of the piercing jet with the cocoon leads to the formation of strong shocks which. The possibility of shock transition strongly depends upon the energy content of the jet. A jet with very high or low values of $E$  and interacting with a moderate strength having cocoon ($m=0.3$) is less affected by it and is not capable of forming the shocks. However, the shock transition takes place for intermediate energies (Figure \ref{lab_vel_m0.3_xi_0.1}). Our hydrodynamic study captures the theoretical picture of the jet collimation by the cocoon in a GRB jet that is repeatedly seen in various numerical studies \cite{nhs14} and the formation of recollimation shocks above the stellar surfaces \cite{lml15}. However, in the current model we are only able to observe a single recollimation shock compared to such multiple shocks seen in some simulations. It is due to complicated structure and the time evolution of the cocoon interaction with the jet.
\item The compression ratio $\cal R$, as well as, the transition location $r_{sh}$ of the generated shock are sensitive to all the free parameters, cocoon strength $m$, jet energy parameter $E$ and the fluid composition $\xi$. Stronger shocks are formed by higher collimation, as well as, the jets with greater energy. 
\item We show that the jets injected at sub relativistic speeds can be driven up to relativistic Lorentz factors following thermally driving. These jets comfortably achieve Lorentz factors $\sim$ few $\times$ $10$. As we have ignored other possible accelerating agents such as magnetic driving and radiation in this study, the Lorentz factors are mildly relativistic. The Lorentz factor of GRB 170817A is constrained to be $\gamma=13.4^{+9.8}_{-5.5}$ \cite{zwm17}. Further, the observationally constrained minimum Lorentz factors for several GRBs with early time radio emission are found to be in a typical range of $5.8-21$ (table 4 of \cite{avs14}) which is again consistent with the magnitudes obtained in this study. The upper limit of Lorentz factors extends by an order of magnitude and the above-mentioned acceleration factors might be responsible there.
%\item In the presence of magnetic fields across the shock, the recollimation shocks generated in this papers would accelerated nonthermal electrons through diffuse Fermi acceleration \citep{be87}. The spectral index of emitted nonthermal radiation by the accelerated particles is given as \cite{vc17}:
%\be 
%s=-\frac{{\cal R}}{{\cal R}-1}
%\ee
%We plot the photon index ($\beta=s-1$) as a function of $\xi$ in figure \ref{lab_s_index_vs_xi} (left) and $m$ (right). The parameters of these figures are same as in figures \ref{lab_vel_m0.8_xi_1.0} (bottom left) and \ref{lab_R_vs_E} (bottom left) respectively. The range of obtained spectra with photon index $-2.5$ to $-3.9$. A large number of gamma ray bursts have been found to be showing such steep power law spectra at high energies \cite{r98,kpb06}. Hence, we iterate that such power laws can be generated due to shocks induced by the cocoon.
%\item Further, as we see that the emitted spectra get harder as $\xi$ increases, we can indirectly infer that harder prompt phase spectra at high energies indicate towards presence of higher baryon fraction. While concluding this, one needs to be aware that the system has other degeneracy in $m$ and $E$ and this is only an indirect hint which needs to be investigated further.
\end{itemize}
\section{Missing links in this work and future prospects}
\label{sec_future}
Besides the importance of this study in exploring the GRB jet dynamics in general relativistic regime along with the dependence of outcomes on the jet composition, the model is quite simple and doesn't consider the jet cocoon mixing, and its possible effects. Also, we ignored the effects of other possible factors like radiation driving and the effects of large scale magnetic fields on the jet. In the next attempts, we will study the effect of the existing radiation field due to the star and the cocoon and will elaborate on their effects on the jet dynamics. In previous works related to radiation driving of X-ray binary jets and the jets in AGNs, we established that the external radiation fields are capable of inducing recollimation shocks in the jets. It is worth investigating to seek such effects in GRB jets. In the presence of magnetic fields across the shock, the recollimation shocks generated in this paper would accelerate nonthermal electrons through diffuse Fermi acceleration \citep{be87}. However, to account for the precise estimation of particle acceleration, one needs to incorporate the effect of the magnetic field on the shock conditions as well as on the jet propagation. Furthermore, the effect of the jet composition on the shock properties in this model has no numerical analogue to the best of our knowledge and can be tested in future simulations to reveal their significance as well their time-dependent nature.
%\begin{figure}[H]
%\begin{center}
%\includegraphics[width=6.5 cm]{s_index_vs_xi.eps}
%\includegraphics[width=6.5 cm]{s_index_vs_m.eps}
%%\includegraphics[width=6.5 cm]{E_m0.eps}
%%\includegraphics[width=6.5 cm]{Ent_m0.eps}
%\caption{Photon index $\alpha$ as a function of $\xi$ (left) and $m$ (right) produced by the radiation spectra generated due to diffusive shock acceleration for Figures \ref{lab_vel_m0.8_xi_1.0} (bottom left) and \ref{lab_R_vs_E} (bottom left), respectively.} 
%\label{lab_s_index_vs_xi}
%\end{center}
%\end{figure}  

\acknowledgments{I am thankful to the anonymous reviewers who helped in clarifying various aspects of the study and I am grateful to Asaf Pe'er for an insightful discussion and important suggestions. I further acknowledge the support from Israel government's PBC program and the European Union (EU) via ERC consolidator grant 773062 (O.M.J.)}

\conflictsofinterest{The author declares no conflict of interest.}
\begin{adjustwidth}{-\extralength}{0cm}

\reftitle{References}

\end{adjustwidth}
\end{document}